# Heterogeneous Quantization Regularizes Spiking Neural Network Activity

Roy Moyal*[1,2], Kyrus R. Mama[1,3], Matthew Einhorn[1], Ayon Borthakur[4], Thomas A. Cleland*[1]

[1] Computational Physiology Lab, Department of Psychology, Cornell University, Ithaca, NY 14853, USA
[2] AI for Science Institute, Cornell University, Ithaca, NY 14853, USA
[3] Neurosciences Graduate Program, Stanford University, Stanford, CA 94305, USA
[4] Mehta Family School of Data Science and Artificial Intelligence, IIT Guwahati, Guwahati, Assam 781039, India

*Corresponding authors: rm875@cornell.edu, tac29@cornell.edu

## Abstract

The learning and recognition of object features from unregulated input has been a longstanding challenge for artificial intelligence systems. Brains are adept at learning stable representations given small samples of noisy observations; across sensory modalities, this capacity is aided by a cascade of signal conditioning steps informed by domain knowledge. The olfactory system, in particular, solves a source separation and denoising problem compounded by concentration variability, environmental interference, and unpredictably correlated sensor affinities. To function optimally, its plastic network requires statistically well-behaved input. We present a data-blind neuromorphic signal conditioning strategy whereby analog data are normalized and quantized into spike phase representations. Input is delivered to a column of duplicated spiking principal neurons via heterogeneous synaptic weights; this regularizes layer utilization, yoking total activity to the network's operating range and rendering internal representations robust to uncontrolled open-set stimulus variance. We extend this mechanism by adding a data-aware calibration step whereby the range and density of the quantization weights adapt to accumulated input statistics, optimizing resource utilization by balancing activity regularization and information retention.

## Introduction

Learning relevant features from unpredictable input distributions is a critical task for any neural network architecture. Under realistic conditions ("in the wild" [1]), robust representation learning hinges on the quality of signal preprocessing, which in turn depends on the successful application of modality-specific domain knowledge. Biological systems excel at solving such ill-posed problems; they embed relevant domain expertise in the functional architecture of their circuitry, deploying computational features including specialized neuron types, synaptic weight distributions, and plasticity rules towards solving specific problems. Neuromorphic computing strategies abide by this principle, treating network architecture as part of the algorithm and favoring efficiency and robustness over generality. What, then, is required of a signal conditioning mechanism in a neuromorphic artificial olfactory system intended for deployment in the wild?

Unregulated signals are extremely challenging to sample and interpret reliably. Naturalistic stimuli, in particular, are prone to source mixing and environmental interference; moreover, they often exhibit intensity ranges spanning several orders of magnitude, with potentially relevant information embedded across multiple scales. The selectivities of primary sensors, the overlaps among their receptive fields, and the limitations of their transduction mechanisms all compound the problem, ultimately imposing bounds on the resolution and discrimination capacity of the sensory system. These considerations are inherent to the problem space and apply to biological and artificial sensors alike.

In olfaction, to achieve specific analyte recognition, one strategy is to narrow the selectivity of a single sensor so that it detects only the analyte of interest. This can be achieved directly for certain chemically distinctive analytes (e.g., oxygen, ammonia). Alternatively, the deployment environment can be controlled such that the sensor is only ever exposed to one analyte to which it is sensitive, thereby rendering the sensor highly selective within the context of that highly regulated environment. This is how synaptic neurotransmitter specificity is achieved within biological brains [2], but is an ineffective strategy in the wild. A third strategy, utilized in biological olfactory systems as well as in biomimetic artificial systems, is to deploy arrays of chemical sensors, each of which may interact with multiple analytes, and achieve specificity by interpreting the patterns of responses across the elements of the sensor array [2–4]. Specifically, response patterns can be correlated across analytes or sensors, owing to similarities in analyte molecular structure or physicochemical properties (depending on the physics of sensor transduction); it is these correlation patterns from which specificity is ascertained. However, these correlations are highly vulnerable to the source mixing, interference, and scale problems presented by unregulated sensory environments. Chemical analytes can be encountered at operationally relevant concentrations ranging from a few parts per billion up to overwhelming intensities that saturate the responses of primary sensors. The sensitivities of different chemosensors to particular elements of a chemical environment can vary by orders of magnitude, and their responses to increasing concentrations can be sharply nonlinear. Different types of chemosensors are vulnerable to different subsets of these problems.

We have previously presented a neuromorphic algorithm, based on the architecture and function of the mammalian olfactory bulb external plexiform layer (EPL; Fig. 1) and implemented on Intel Loihi. This method learns multiple arbitrary representations rapidly and sequentially, without catastrophic forgetting, and subsequently can identify them even in the presence of competitive background interference [5]. Specifically, that EPL network is robust to impulse noise affecting a subset of its sensors and to a limited amount of variance across all sensors, but it relies on input statistics that are fundamentally well-behaved. Transforming the uncontrolled variance of the external world into a sufficiently regularized, yet still informative, activity distribution must occur prior to EPL activation. In the biological olfactory system, there are multiple mechanisms described that preserve representational integrity across wide analyte concentration ranges [6] and compress it into the limited dynamical range of olfactory bulb circuitry [7]. It also is established that diffuse lateral shunting inhibition in the olfactory bulb glomerular layer (GlomL; Fig. 1) delivers divisive global normalization across the array of olfactory bulb *columns*, each of which is associated with one of its hundreds of different receptor types. This enables concentration invariance in downstream areas by ensuring that the total activity is stable across input samples [8–11]. However, we here show that this normalization, while necessary, is not sufficient to provide the EPL with statistically well-behaved input.

To resolve this problem, we demonstrate a mechanism predicated on signal duplication and heterogeneity within columns (gain diversification) that implements input regularization at the point of quantization, in which the analog responses of external tufted (ET) cells in the GlomL are transduced into the temporally patterned spiking activity of mitral cell principal neurons (MCs) in the EPL (Fig. 1). By this method, unpredictable and unregulated sensory input can be transduced and regularized for effective further processing by downstream spiking layers, without foreknowledge or the need to operationally tune hyperparameters. We further describe and optimize this method with respect to activity regularization, information retention, and resource utilization.

## Network Architecture

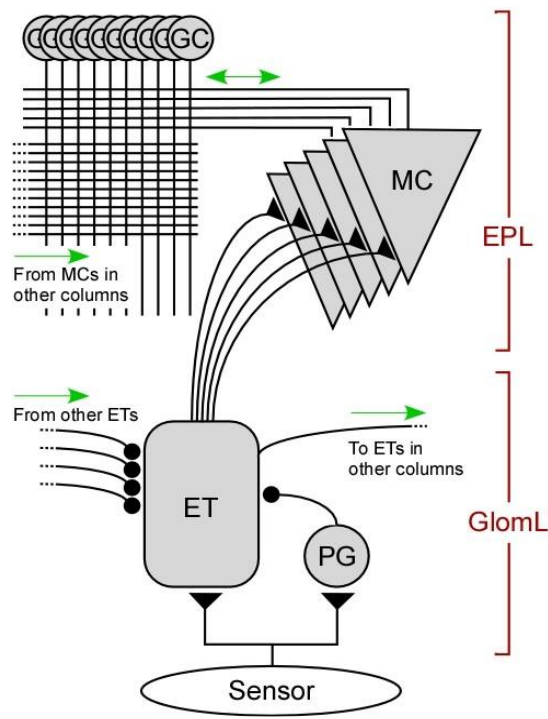

**Figure 1**. **Neuromorphic network diagram of a single sensor-associated column**. The design is directly derived from the circuit architecture of one column of the mammalian olfactory bulb (OB). Input from a specific sensor class is delivered to external tufted (ET) neurons in the first layer, based on the glomerular layer (GlomL) of the olfactory bulb. In the artificial network, ET cells are non-spiking and deliver graded feedforward synaptic excitation onto mitral cell (MC) principal neurons, generally with heterogeneous synaptic weights. ET cells also inhibit one another across columns in a nonspecific mutually inhibitory network that subsumes the role of inhibitory ET cell-driven periglomerular (PGe) and superficial short axon (sSA) cells in the biological network [8,10]; this network effects global normalization across the sensory input field. MCs are spiking neurons that initiate dynamical, spike timing-dependent interactions with inhibitory granule cells in a plastic network modeled after the external plexiform layer (EPL) of olfactory bulb [5]. Importantly, MCs excite GCs with a fixed probability across the full network, irrespective of column, whereas GCs inhibit MCs only within their home column. In the present study, GC→MC inhibition was disabled to isolate the effects of the quantization step (ET→MC).

# Results

## Dataset Properties and Signal Conditioning

The response signatures of chemosensor arrays to analytes of interest and their environmental backgrounds are often not statistically well-behaved. High quality signatures are high-dimensional, providing diversity across sensors both (1) in the levels of activation exhibited (i.e., some strongly activated, some weakly or not at all) and (2) in the *channel coding* sense, in which different analytes activate correspondingly different complements of sensors. Well-behaved inputs exhibit the former diversity in a consistent distribution, irrespective of which specific sensors are strongly or weakly activated.

Here, we sought to transform a wide variety of raw input signatures to a common, well-behaved spike phase distribution with minimal information loss. To that end, we generated two synthetic datasets, each comprising eight analytes and activating an array of eight artificial chemosensors. In the first, the *concentration* dataset (Fig. 2A-D, S1), sensor receptive fields were balanced, but were distinguished by different overall sensor response levels—possibly as a result of varying analyte concentrations, or else corresponding to a range of analytes among which that sensor array was highly sensitive to some but not others. In the second, the *saturation* dataset (Fig. 2E-H, S1), the analytes were poorly encoded by the primary sensor transduction process, with some or most of the sensors saturated by analyte presentation, and with relatively few sensors exhibiting intermediate activation values that are sensitive to small differences in analyte quality. Because we here emphasize the problem of data regularization—i.e., transforming input distributions to accommodate the operating range of a downstream spiking network—we depict activity distributions sorted along the abscissa by amplitude for clarity, de-emphasizing channel coding differences among analytes (unsorted activation profiles are depicted in Fig. S1 and in matrix colormap form in Fig. 2D, 2H). To render the problem more difficult, different samples of the same artificial analyte were jittered by adding a small amount of Gaussian noise (see *Methods*).

Raw sensor outputs were each delivered to a corresponding analog neuron modeled after the external tufted (ET) cells of the olfactory bulb (Fig. 1). Initial normalization was performed for each of the eight analytes presented by globally scaling all ET cell activation levels such that their activity summed to a constant (Fig. 2B, 2F). This corresponds to the biological architecture, in which ET cells interact via inhibitory interneurons in a broad, nonspecific lateral network, thereby normalizing chemosensory input into a relational representation [7,9,10]. In a neuromorphic instantiation, unconstrained by Dale's principle, the ET cells simply inhibit one another directly in an all-to-all lateral network while simultaneously exciting MC principal neurons, as depicted in Fig. 1. Notably, even after normalization, the distribution of activity levels across the eight ET cells differed as a function of the analyte presented (Fig. 2C, 2G); for example, analyte 1 from the *concentration* dataset evoked a narrow range of moderate activation levels across columns, compared to analyte 8 which evoked close to zero activity in some ET cells and near-maximal activity in others.

The normalized representations generated in the analog ET cells activated MC principal neurons within the same column. MCs are spiking neurons; however, rather than representing their activation level with a mean spike rate, a sinusoidal background rhythm is applied (known as the *gamma oscillation* after its biological equivalent). Spike rates are constrained to zero or one spikes per gamma cycle, and the activation level of a given MC is represented by the relative phase of its spike. This phase coding enables a single spike to embed a numerical value of arbitrary precision, works naturally with established online learning rules such as spike time-dependent plasticity (STDP) [12], and underlies a previously described iterative signal restoration process in a neuromorphic EPL circuit [5]. The synchronization of MC spikes governs the recruitment of granule cell interneurons into the activated ensemble (GCs; Fig. 1), as previously described [5] (in the present study, GC→MC recurrent inhibition was disabled, to isolate the quantization effect and assess the initial recruitment of MCs and GCs into the active representation). Here, we establish a fixed precision of 50 discrete phases per cycle. This temporal resolution can be adjusted as required, potentially improving precision or memory capacity at the cost of higher energy-to-solution.

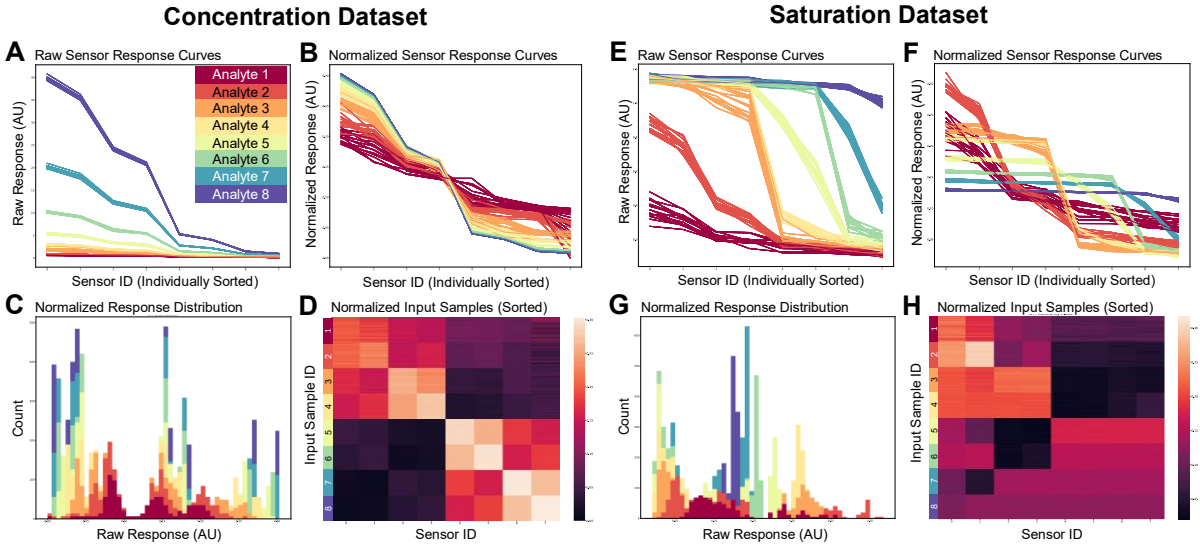

**Figure 2**. **Raw and normalized sensor responses to test samples from two synthetically generated datasets.** Samples, accumulated over eight cross validation folds (train-test splits), are color coded by analyte class. (**A**) Raw sensory dataset with balanced sensor affinities and systematic differences in synthetic analyte concentration. Response curves are sorted individually from most to least active, to emphasize differences in slope (see Fig. S1 for unsorted data). (**B**) Dataset from A following global normalization. (**C**) Stacked histograms depicting normalized synthetic sensor response distributions (binned counts). (**D**) Normalized sensor responses in matrix form, sorted row-wise by analyte class to emphasize similarity structure, with sensors on the horizontal axis and samples on the vertical axis. (**E-H**) A sigmoid transformed variant of the first dataset, with imbalanced response magnitudes and systematic differences in sensor saturation across analytes. In both datasets, a small amount of Gaussian noise was added to render the classification problem more difficult (see *Supplementary Methods*).

**Heterogeneous Quantization Regularizes Spiking Activity and Mitigates Lossiness**

Biological systems deploy diverse mechanisms to accommodate natural inputs with wide dynamic ranges. Among these strategies, heterogeneity (e.g., of connection topologies or synaptic weights) is prominently employed across sensory systems [6,13–15]. In the mammalian olfactory bulb, tens of principal neurons are associated with each column, inheriting afferent input from the same primary sensory receptor class. We tested whether implementing heterogeneity across these duplicated parallel pathways—multiple MCs per column (Fig. 1)—could help regularize layer utilization at the quantization step without resorting to foreknowledge of environmental statistics or a time-consuming hyperparameter optimization procedure. Specifically, we studied the effects of heterogeneity among ET→MC synaptic weights and MC spike thresholds (see *Supplementary Data*); we here focus on the former as it better segregates this transformation from subsequent computations within EPL [5].

To obtain a set of baseline measures, we first trained networks with *homogeneous* ET→MC weights (and uniform MC thresholds) on pure samples of simulated analytes from our two datasets. Network sizes varied, reflecting an MC column *duplication factor* of 4x, 8x, 16x, or 32x in each of the eight columns, but ET→MC weights (gain levels) were all identical. The network was tested with 10 possible weight values, drawn from a $\frac{1}{x}$ distribution (see *Supplementary Data*) and covering the input range $[.01, 1]$ across sensors. Training was two-shot (i.e., 16 samples in total, two from each analyte class), and eight-fold cross validation was used to obtain confidence intervals around the measures we report (see *Methods*).

Fig. 3A depicts MC spike phase responses as gain levels (*W*) are increased, illustrating that no single weight can generate a high-quality representation of all tested analytes. To quantify this, we measured the mean analyte classification accuracy for each dataset across ten fixed ET→MC synaptic weights. We trained separate support vector machines (SVM) with a radial basis function (RBF) kernel (see *Methods*) on MC and GC phase response vectors as well as on ET activation levels (equal to the normalized analog input vectors, serving as a baseline). In the concentration dataset, information was lost only at the lowest and highest weights, whereas in the saturation dataset, information was fully retained, on average, only within a narrower weight range (Fig. 3B).

# Homogeneous Quantization

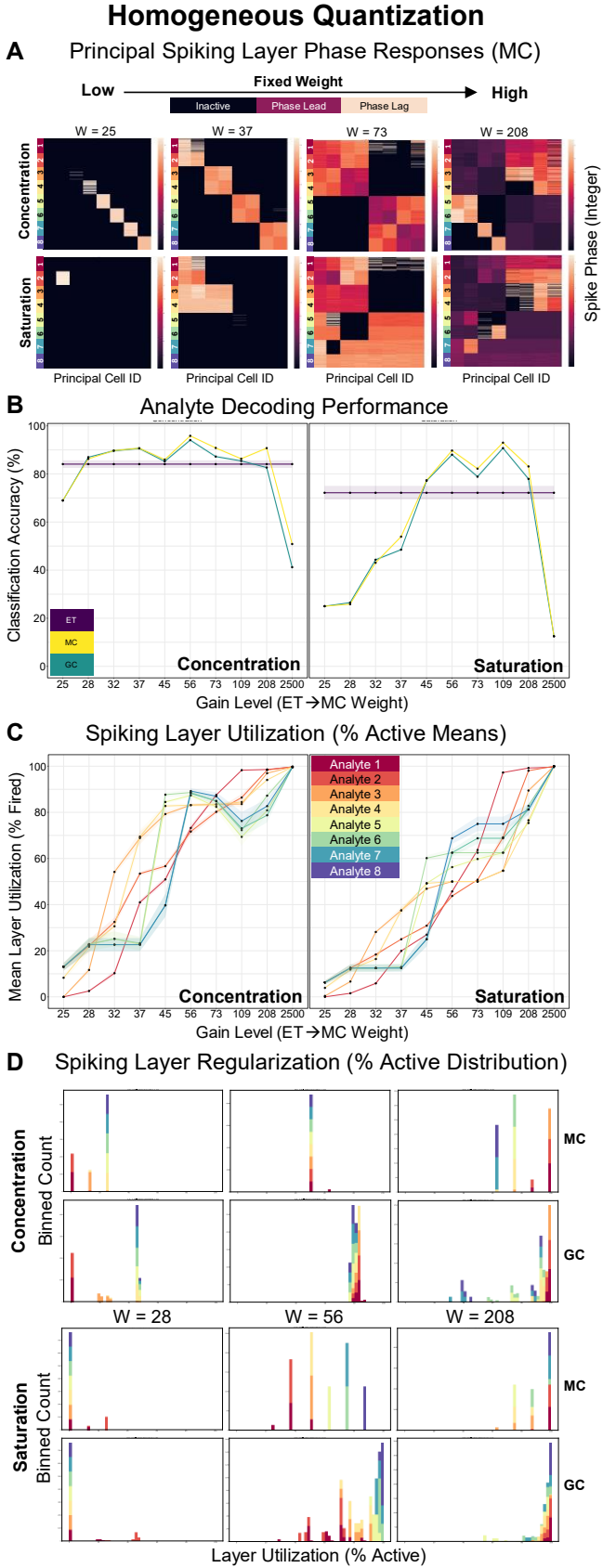

**A** Principal Spiking Layer Phase Responses (MC)

**B** Analyte Decoding Performance

**C** Spiking Layer Utilization (% Active Means)

**D** Spiking Layer Regularization (% Active Distribution)

Increasing ET→MC gain also broadened the activation of MCs (Fig. 3B). Examining the analyte-specific recruitment of MCs across these different weights also revealed that, even at moderate values, some analytes activated nearly the full complement of MCs (Fig. 2B, 2F)—an undesirable state even if reliable phase differences among activated MCs were maintained. Finally, and most critically, different analytes recruited widely different numbers of MCs and GCs from the same parameterized network (Fig. 3D). This high variance in neuronal recruitment disrupts the balance between excitation and inhibition required for robust neuromorphic network operation; if the network was well parameterized for some analytes, it necessarily would be poorly parameterized for others. Overall, no optimal quantization parameters could be found that would consistently preserve similarity structure, retain analyte decodability, and ensure the stability of EPL network dynamics and plasticity across analytes and datasets.

**Figure 3. Outcome of homogeneous quantization.** A fixed ET→MC weight value is adjusted as a control parameter on different simulation runs. Possible weights are distributed $\frac{1}{x}$ to emulate a set of equidistant MC thresholds covering the input range $[.01, 1]$ (see *Supplementary Data*). Results are aggregated over eight cross validation folds. (**A**) Principal layer (MC) spike phase responses to test samples from the two datasets. Spike phase responses range across $[1, 50]$ and inactive units are coded as zeros (black). Within each matrix, rows correspond to individual samples and are sorted by analyte class. Columns correspond to individual MCs. (**B**) Line plots depicting classification performance of hyperparameter optimized RBF SVMs trained separately on ET (purple), MC (yellow), or GC (teal) phase responses, illustrating the effects of increasing gain levels (ET→MC weight) on information representation in homogeneous networks of varying sizes (4x to 32x; see *Methods*), in the presence of concentration variability and sensor saturation. The analog ET responses function as a baseline, being identical to the normalized input test samples. (**C**) Spiking layer utilization, averaged across models and over the MC and GC layers. The curves depict the mean percentage of active units as gain increases; line colors correspond to analyte class labels. (**D**) Stacked histograms depicting MC and GC layer utilization (%) as gain increases, for the concentration (*top*) and saturation (*bottom*) datasets, at 32x homogeneous duplication. Across panels, ribbons represent the standard error of the mean over eight cross validation folds.

We sought a transformation in which a single parameterized network could encode this full diversity of analytes within a predictable and balanced regime. We therefore repeated the above simulation experiments with *heterogeneous* ET→MC weights, at MC duplication factors of 4x, 8x, 16x, and 32x. The weights were similarly $\frac{1}{x}$ distributed and covered the same input range, but this time the weight variance was deployed across multiple MCs within each column rather than across simulation runs. This yielded more balanced spike phase responses (in Fig. 4A, on the abscissa, MCs with a common ET afferent are adjacent and sorted by weight, low to high). To assess the performance of this strategy, we fit general linear models (GLM) to the dependent variables *information* and *regularization*, with *condition* (homogeneous or heterogeneous)*, duplication factor, layer* (MC or GC), and *dataset* as fixed factors, and *cross validation fold* as a random factor. Unless otherwise stated, general linear contrasts were Tukey-corrected within (but not across) nested levels of the relevant factors.

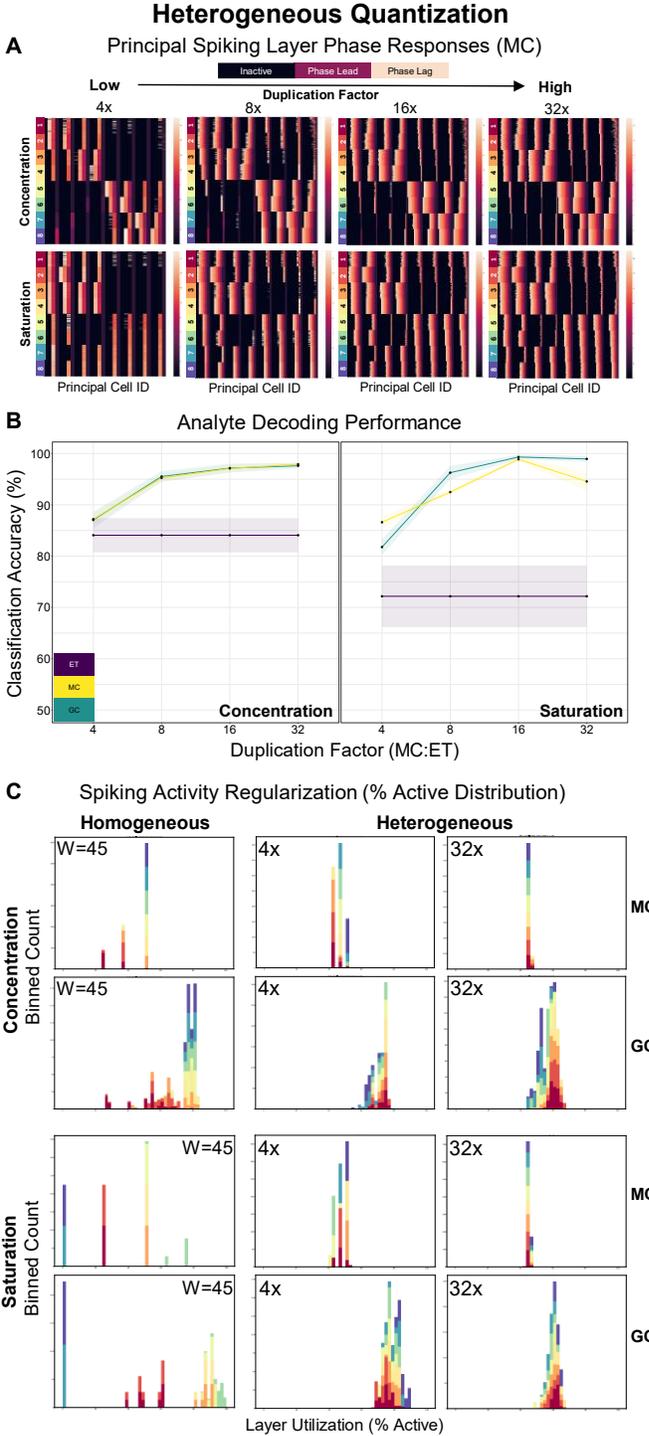

**Figure 4. The effect of heterogeneous quantization on regularization and information.** Performance is examined at four duplication factors: 4x, 8x, 16x, 32x. The weights of synapses originating from the same ET cell are distributed $\frac{1}{x}$ to emulate equidistant MC thresholds covering the input range [.01, 1] within each MC column (see *Supplementary Data*). Results are aggregated over eight cross validation folds; ribbons represent the standard error. (**A**) MC spike phase response matrices at each duplication factor, separated by dataset. (**B**) Line plots depicting classification performance of hyperparameter optimized RBF SVMs trained separately on ET (purple), MC (yellow), or GC (teal) phase responses, demonstrating a marked increase in information across layers as the duplication factor increases. ET responses are equivalent to the normalized input test samples and serve as a baseline. (**C**) Stacked histograms depicting MC and GC layer utilization (%) for two heterogeneous conditions (4x and 32x; *rightmost columns*), alongside their 32x homogeneous counterparts (mid-level gain of *w*=45 depicted, *leftmost column*). Heterogeneous weights yield consistent utilization of the MC and GC layers irrespective of analyte identity. Colors correspond to analyte class labels.

In the model fit to *classification accuracy*, we observed main effects for all factors $F([1,3], 217) > 9.372, p < .0025$. In addition, except for *layer* by *dataset* and *condition* by *layer* by *dataset*, all two-way and higher order interactions were significant, $F([1,3], 217) > 2.702, p < .0465$. Planned contrasts revealed that analyte classification accuracy was higher with heterogeneous than with homogeneous ET→MC weights when either MC responses, $t(217) > 21.773, p < .0001$, or GC responses, $t(217) > 22.268, p < .0001$, were used as input vectors to the classifier (Fig. 4B). In the *regularization* model, all main effects were significant, $F([1,3], 217) > 4.266, p < .006$. Two-way interactions of *condition* by *layer* and *condition* by *dataset* also were observed, $F(1, 217) > 9.739, p < .002$. Specifically, the number of active MCs was more consistent across diverse test samples in the heterogeneous than in the homogeneous models; across duplication levels, the standard deviation of MC and GC utilization was lower (better) when variable ET→MC weights were used, $t(217) > 19.026$, $p < .0001$. These findings indicate that, all else being equal (including network size), a data-blind $\frac{1}{x}$ distribution of ET→MC quantization weights within each column regularizes neuronal activation and enhances information retention in downstream spiking layers (Figs. 4C, S2, S3).

**Adaptive Quantization Jointly Optimizes Resolution and Regularization**

A $\frac{1}{x}$ distribution of ET→MC weights within each column (or a uniform distribution of MC thresholds; see *Supplementary Data*) provides data-blind optimal regularization of unregulated sensory inputs, thereby presenting statistically well-behaved inputs to the spiking network of the EPL (Fig. 1); higher duplication factors generate correspondingly more reliable distributions (Figs. 4C, S2). This flat prior is generally responsive to all possible sensory input profiles. However, to the extent that the statistics of a test environment are predictable, a data-aware modification of this approach can maintain or improve performance while expending fewer resources (i.e., achieving a given degree of information retention with a lower duplication factor). If the full potential activation range of a given sensor (here, $[0, 1]$) is not predicted to be used within a test environment, mapping only the range of expected activation levels will increase resolution at the cost of returning a low-quality representation should this range be exceeded. Second, the likelihood distribution of particular activation levels of each sensor within a given environment can be learned, and the distribution of the quantization weights mapped accordingly, assigning higher resolution to the most probable or most task-important ranges of sensor activation.

This is a deeply biological design principle. Brains, over behavioral, developmental, and evolutionary timescales, adapt to the statistics of their input and exploit regularities to optimize processing. This manifests in functional architectures that vary widely across sensory systems. Moreover, biological circuits conserve resources by allocating them selectively, in both a structural (e.g., a greater density of cone photoreceptors in the fovea than at the periphery [16,17]) and a functional sense (e.g., saccading behavior [18]). In many artificial systems, it is desirable to allocate bandwidth on demand; here, we leverage this design principle to minimize resource allocation while jointly optimizing information retention and activity regularization.

In this final set of simulation runs, we introduced an adaptive pre-training calibration step, utilizing the same small hold out set used for two-shot training. ET→MC weights, which were uniformly spaced ($\frac{1}{x}$) in the standard heterogeneous condition, were allowed to vary in range and density (independently across MC columns) to match the distribution of the calibration measurements; in other words, we used the calibration set to construct a non-flat prior on the input distribution. We tested two variants of the calibration procedure: in the *scaled* condition, weights covered the range between the maximal and minimal observation recorded by each sensor in a $\frac{1}{x}$ distribution; in the fully *adaptive* condition, weights were piecewise interpolated to reflect the calibration set distribution, and the number of segments (dictating the smoothness of the interpolation) was identical to the duplication factor (see *Methods*). We hypothesized that adaptation would enhance the system's capacity for discrimination at lower duplication factors while sacrificing some degree of regularization (i.e., the standard deviation of layer utilization would be somewhere in between the homogeneous baseline and the generic heterogeneous condition).

As before, we fit GLMs to our *information* and *regularization* measures, with *condition* (this time with four levels: homogeneous, heterogeneous uniform, scaled, and adaptive), *duplication*, *layer* (MC or GC), and *dataset* as fixed factors, as well as *cross validation fold* as a random factor.

In the model fit to classification accuracy (Fig. 5A), we observed main effects for every factor, $F([1,3], 441) > 5.805$, $p < .0164$. Except *layer* X *dataset* and all three-way interactions involving *layer*, every interaction effect was significant, $F([1,3,9], 441) > 2.719$, $p < .0273$. Across duplication factors and layers, all heterogeneous condition variants yielded higher classification accuracy compared to their homogeneous counterparts, $t(441) > 18.254$, $p < .0001$. At the lowest duplication level, 4x, the scaled and adaptive variants yielded boosted accuracy relative to the uniform heterogeneous condition, $t(441) > 5.188$, $p < .0001$, illustrating the compensatory effect of data-aware calibration.

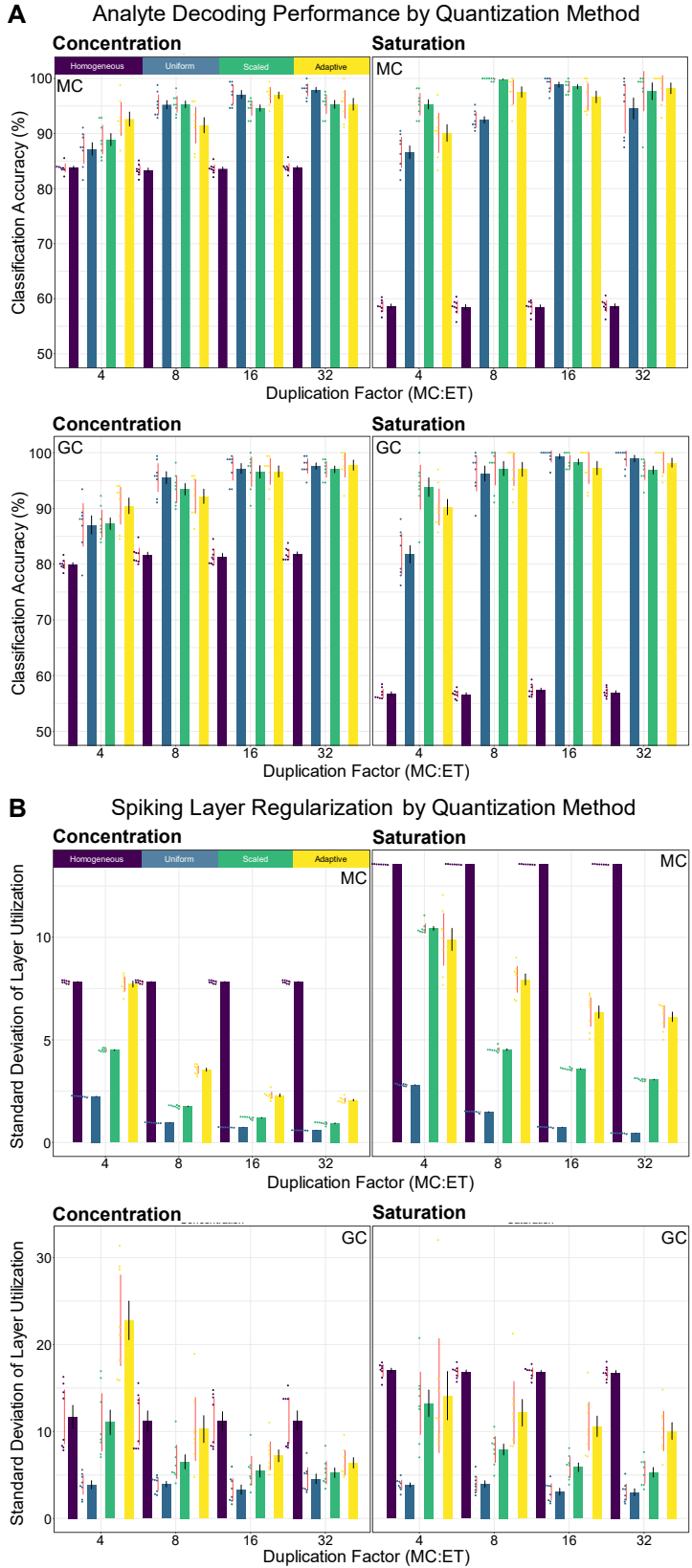

At 8x duplication, only MCs showed an advantage for the scaled compared to the uniform heterogeneous condition, $t(441) = 4.32$, $p = .0001$. By 16x and 32x, all heterogeneous networks had reached ceiling performance, yielding no differences among the uniform, scaled, and adaptive variants, $t(441) < 1.643$, $p > .3556$. In the heterogeneous conditions (but, critically, not in the homogeneous baselines; $p > .6941$), increasing duplication from 4x to 32x provided a considerable boost to information retention; for both MCs and GCs, classification accuracy increased monotonically with duplication factor. The largest difference was observed between 4x and 8x across conditions, $t(441) > 3.675$, $p < .0007$, followed by 8x and 16x in all conditions except *scaled*, $t(441) > 2.508$, $p < .0347$. By the final step, 16x to 32x, an asymptote had been reached across conditions and layers, resulting in null differences, $t(441) < 2.032$, $p > .1121$.

**Figure 5**. **Comparison of quantization methods**. Calibration optimizes resource utilization and trades off regularization for enhanced resolution in ranges where observations are expected to be found. In the *scaled* and *adaptive* conditions, Weights encode a non-uniform prior on the input distribution; weights are interpolated along segments whose boundaries separate cluster centers on the line (Jenks natural breaks, given the calibration observations for each sensor [19]). The number of segments is controlled by a smoothing factor. In the *scaled* condition, a single segment is used, bounded individually for each sensor by the minimal and maximal values recorded. In the *adaptive* condition, multiple segments are used, yielding a better fit to the calibration data. (**A**) Classification performance of cross validated and hyperparameter optimized RBF SVMs, trained separately on MC and GC phase responses from individual models with different MC duplication factors. (**B**) Utilization regularization (standard deviation of percentages of active MC and GC neurons), illustrating the regularization-information tradeoff as a function of the MC duplication factor. In all panels, bars depict mean classification accuracies within each layer, condition, and level of duplication factor. Black ticks represent the standard error of the mean. Red vertical lines correspond to 95% confidence intervals around each mean. Scattered points are individual observations from one of eight cross validation folds.

In the utilization regularization model (Fig. 5B), we found main effects of all factors, $F([1,3], 441) > 91.858$, $p < .0001$. With the exception of *duplication* by *dataset*, $F(9, 441) = 2.041$, $p = .1074$, all interactions were significant, $F([1,3,9], 441) > 2.721, p < .0059$. General linear contrasts revealed that, across spiking layers, utilization regularization was worse (more variability) in the homogeneous condition relative to the three heterogeneous conditions, $t(441) > 2.625, p < .0442$, with one exception; for GCs, at 4x duplication, the adaptive condition was worse than the baseline, $t(441) = 5.735, p < .0001$. For GCs, utilization variability was higher in the scaled and adaptive conditions relative to the uniform heterogeneous condition at duplication levels 4x, 8x, and 16x, $t(441) > 3.77, p < .0001$. By 32x, utilization was similarly regularized in the uniform and scaled variants, $t(441) = 2.101, p = .1544$, and both showed an advantage relative to the fully adaptive variant, $t(441) > 4.104, p < .0003$. For MCs, worse regularization was observed in the adaptive compared to the uniform variant across duplication factors, $t(441) > 4.995, p < .0001$. That was also true for the scaled relative to the uniform condition at 4x and 8x, $t(441) > 2.677, p < .0385$, and for the adaptive relative to the scaled condition at all duplication levels, $t(441) > 2.697, p < .0364$, except 4x, $t(441) = 1.872, p = .2416$, consistent with the existence of a regularization-resolution tradeoff.

The above outcomes demonstrate that the benefits of heterogeneous quantization at modest resource multipliers can be enhanced by adopting a data-aware quantization strategy. With lower duplication, modifying the range and distribution of ET→MC weights on a sensor-by-sensor basis, based on calibration observations, significantly boosted information retention with a small negative impact on spiking activity regularization. This property can be useful with denser sensor arrays, where using higher duplication factors can become resource intensive.

## Discussion

The goal of odor source separation and identification from real-world data presents a challenging problem. Whereas some simple tasks can be performed by analyte-specific electrochemical sensors (e.g., $CO_2$ detectors) that exhibit negligible sensitivity to background sources [20], most chemosensors are responsive to broad ranges of chemical analytes with related properties. Accordingly, biological and most artificial chemosensory systems pursue the richer strategy of deploying arrays of partially selective sensors; these chemosensor arrays achieve response specificity in combination, and additionally enable the assessment of similarity between sources. Whereas both individual odors of potential interest and multisource odor scenes constitute linear combinations of analytes present at different concentrations, the mixing of these analytes can exert nonlinear and even nonmonotonic effects on cross-responsive chemosensors [2], which effectively occlude diagnostic activity patterns across the array.

Neuromorphic algorithms, inspired by specific computational strategies of the mammalian olfactory system, have been trained to rapidly learn and reconstruct arbitrary odor source signatures from chemosensor array data in the presence of background interference [5]. However, such networks perform best when tuned to the statistics of well-behaved inputs, normalized and predictable in their activity distributions. Deployment of chemosensor arrays in the wild exposes these networks to additional, powerfully disruptive effects, including (1) high variance in analyte concentrations, which is nonlinear and dwarfs the quality-based signal variance upon which source identification depends, and (2) odor environments that range from sparse to richly complex, yielding varied distributions of activity levels across different sensors in the array even after normalization. Regularizing spiking activity distributions while minimizing lossiness is a prerequisite for more sophisticated sensory constructions, such as hierarchical category learning [21].

To address these challenges, we expanded and unified our previous neuromorphic models of the olfactory bulb glomerular layer [1,22,23] and of the EPL [5], the latter leveraging gamma-like inhibitory oscillations [24–27] to induce a spike-phase code and partition subsequent stimulus processing into cycles. Like the biological system that inspired it, the present network routes sensor outputs to ET cells [28] within a glomerular layer circuit that normalizes activity by lateral shunting inhibition (division) [10]. Here, we investigated the computational properties and effects of the quantization step, where the normalized analog representations of the ET cell layer are transformed into the spike phase vectors of the MC layer. The choice of analog-to-spiking synaptic weights is critical, as it constrains the dynamic range and resolution of downstream representations. Our goal was to jointly optimize information (discriminability), spiking activity regularization, and resource utilization.

We deployed a spectrum of *heterogeneous quantization* strategies, all involving the use of variable ET→MC weights; this mechanism amounts to gain diversification among parallel outputs (as in spare receptor theory [6]), enabling downstream layers to represent signals spanning a wide dynamic range while avoiding a combinatorial explosion of resources. The quantization weights can be conceived of as a prior on the input distribution. In the *uniform* variant of the heterogeneous condition, the weights were $\frac{1}{x}$ distributed (equivalent to linearly spaced thresholds; see *Supplementary Data*) and covered the full range of normalized inputs within each MC column, corresponding to a flat prior on each sensor's response profile. This yielded optimal spiking activity regularization across test samples and a marked improvement in the information decodable from MC and GC spike phase responses compared to the homogeneous baseline.

While the generality of the flat prior achieves the primary goal of enabling a single parameterized network to be well adapted despite the unpredictability of open-set sensor deployment in the wild, it also is true that adapting to the statistics of the chemosensory environment can yield superior performance at lower cost, particularly with smaller duplication factors. To characterize this option, we calibrated the distribution of the ET→MC quantization weights to the environment; specifically, the weights were allocated more densely to ranges in which data were more likely to be found, based on the individual sensor responses observed in a pre-training calibration step. For richly structured datasets, such as those used in the present study, this approach can improve fine discrimination capacity without excessive resource scaling and with a relatively small impact to regularization metrics. This tuning method amounts to piecewise interpolation and allows the user, by adjusting the number of segments (i.e., a smoothing factor), to regulate bias and variance in the model.

Learning sensory representations online, under natural conditions, involves a quantization problem whose outcomes are sensitive to gain modulation. This is especially true in the presence of large sensor arrays, multiscale structure, and weakly diagnostic features. The choice of quantization weights, implicit in many models, determines what orders of magnitude can be represented in the spiking activity of downstream layers. Since the relevant scales are typically not known *a priori*, the deployment of heterogeneous parallel paths—whether data-blind or data-aware—can provide principled solutions with quantifiable bounds. Heterogeneous quantization is a viable alternative to intensive hyperparameter grid searches and is particularly suited to intelligent edge devices.

## Methods

### Synthetic Data Generation

Artificial input vectors were generated that emulate the dynamic range and saturation problems commonly observed in artificial chemical sensors deployed in the wild [29,30] and in the statistics of olfactory sensory neuron (OSN) responses.

*Affinity Matrix.* We designed an affinity matrix $G_{8 \times 8}$ (sensor by source) defining the similarity structure of an artificial odorant space. Values were picked so as to induce a balanced, three-level category hierarchy among eight simulated analyte classes. Receptive fields were distributed symmetrically, such that each analyte had one sensor highly sensitive to it, while others (primarily responsive to other sources) were weakly to moderately sensitive to it (see *Supplementary Methods*). The sparsity of the affinity matrix was defined as the Hoyer normalized sparsity measure, $\frac{1}{\sqrt{N}-1}(\sqrt{N} - \frac{\ell_1}{\ell_2})$ [31,32], applied to each sensor's affinity vector. The mean sparsity of the affinity matrix $G$ was fine tuned by iterative element-wise exponentiation followed by row-wise $l_1$ normalization, $\forall_{0 \leq i < n} \, G_i \leftarrow \frac{G_i^\alpha}{||G_i^\alpha||_1}$, until an overall Hoyer measure of 0.4 was arrived at, emulating moderately selective artificial sensors. This transformation enables the design of hierarchical discrimination problems with a high degree of control over synthetic sensor properties, receptive field distributions, and source mixing proportions.

*Concentration Dataset.* To generate a dataset containing samples with large systematic differences in dynamic range, we multiplied the affinity matrix $G$ by eight-element sample vectors corresponding to pure artificial odorants whose concentrations were exponentially distributed with analyte-characteristic means. To render the classification problem more difficult, we introduced additional variance among samples by injecting a small amount of Gaussian noise to the responses, centered at zero and varying in scale across sensors (for exact parameterization, see *Supplementary Methods*).

*Saturation Dataset.* The total output emitted by a biological OSN unit can be thought of as the sum of multiple sigmoid activation functions with a scaling parameter $s$ and a translation parameter $t$ [6]. This can yield systematic sensor saturation effects under some conditions. We generated an additional dataset modeling this problem by applying a sigmoid transform to the concentration set samples (for exact parameterization, see *Supplementary Methods*). For both datasets, the noise standard deviation was selected so as to render the classification problem nontrivial but tractable. Other free parameters were chosen to isolate and exaggerate the geometric properties of interest.

**Network Architecture**

Network structure followed the circuit motifs of the mammalian olfactory bulb [8], extending and modifying past iterations of our *Sapinet* model [1,5,22]. The present network combines a spiking representation learning module, corresponding to the bulbar *external plexiform layer* (EPL), with an analog signal conditioning module inspired by the *glomerular layer* (GlomL).

*Neurons and Connection Topology*

All spiking units in the model were leaky integrate-and-fire (LIF) neurons, integrated using the fourth order Runge-Kutta method with an update time constant of $1\ ms$. The evolution of the membrane potential was governed by the dynamical equation $\tau_m \frac{dv}{dt} = \frac{I}{g_L} - (v - E)$, where $v$ is the membrane potential, $\tau_m$ the membrane time constant, $I$ the input current, $g_L$ the leak conductance, and $E$ the resting potential. Membrane potential was reset to $E$ upon exceeding a voltage threshold $v_{th}$.

The signal conditioning portion of the network, tasked with normalization and quantization, consisted entirely of analog neurons (for an input current $I$, $\frac{dv}{dt} = I$). Olfactory sensory neurons (OSN) represented sensor responses and directly corresponded to the elements of the input vectors. 8 OSNs projected to 8 external tufted (ET) cells in a one-to-one pattern. PG cells, depicted in Fig. 1, were omitted from the present analysis for simplicity.

Mitral cells (MCs) constituted the principal excitatory spiking layer. They followed a columnar organization principle, where each MC within a column received its primary external input from one ET cell, corresponding to a unique artificial sensor (OSN). All MCs were identically parameterized, $\tau_m = 4$, $g_L = 1$, $E = -60$, $v_{th} = -55$ (fixed).

Depending on model parameterization, there could be several MCs per column, enabling gain diversification through weight or threshold heterogeneity. In this experiment, the ratio of MCs to ETs (*duplication factor*) could be fixed at 1, 4, 8, 16, or 32 (yielding 8, 32, 64, 128, or 256 MCs, respectively). In some cases, ET→MC synapses within the same column varied in weight values, spanning different ranges in different experimental conditions (for details, see *Experimental Design*).

Granule cells (GCs) comprised a second spiking layer. Across conditions, there were four times as many GCs as there were MCs, resulting in a shallow and wide architecture conducive to the learning of basic chemosensory features. All GCs were identically parameterized, $\tau_m = 4$, $g_L = 1$, $E = -60$, $v_{th} = -50$ (fixed). GCs received random input projections from the MC layer, with MC→GC connection probability kept proportional to the ratio of MCs to ETs, $p = \frac{1}{MC/ET}$.

*Oscillations and Temporal Dynamics*

MCs received, at all times, additive sinusoidal input in the range $[-40, 0]$ with a period of $50\ ms$ (simulation steps), segmenting the activity and allowing for iterative convergence of spike-phase representations to an attractor [5]. Accordingly, all spiking unit voltages and refractory periods were reset every $50\ ms$. By enforcing a refractory period, all spiking units were forced to fire only once within each cycle of the background oscillation.

This mechanism follows the example of gamma oscillations in the olfactory bulb [27], by which periodic inhibition synchronizes spiking activity across columns [24–26]. In this mechanism, alternating spiking and quiescent periods enable a within-cycle spike precedence code to form, in which earlier spike phases correspond to stronger inputs.

*Synapse Models and Learning Rules*

In the GlomL, connections between OSNs and their respective ETs were simple relays ($w = 1$). ET cells were fully laterally connected with shunting synapses (including self-connections); this divisive inhibition implemented $l_1$ normalization, ensuring $ET(x) = \frac{x}{||x||_1}$ for any input vector $x$.

The ET→MC quantization synapse weight distributions varied across experimental conditions. In the uniform heterogeneous condition, $\frac{1}{x}$ distributed weights were chosen to cover the normalized input range $[.01, 1]$ (that is, to allow at least one MC to fire for an input at the network's detection limit, here set to .01, subject to LIF parameterization). In others, weights could cover more limited input ranges or vary in density (see *Experimental Design*).

MC→GC synapses were additive. Their weights were initialized to $w = 25$ and updated during training in accordance with a standard Hebbian STDP rule, $\Delta w_{ij} = \alpha_+ \cdot e^{\frac{t_j - t_i}{\alpha_+}}$ for potentiation, $\Delta w_{ij} = -\alpha_- \cdot e^{\frac{t_i - t_j}{\alpha_-}}$ for depression, where $\tau_+, \tau_- = 4, \alpha_+ = .3125, \alpha_- = 1.25$ [5,12]. Updates were performed on every cycle of the background oscillation, as spiking neurons were guaranteed to fire at most once. Learned weights were clamped to the range $[0, 30]$, and a half-cycle synaptic delay ($25\ ms$) ensured GCs could only fire when MCs were quiescent.

**Experimental Design**

We tested the effects of the quantization transformation (from analog ET responses to MC spike phase representations) on the following dependent variables: (1) variability in spiking layer utilization (*regularization*); and (2) the quality of the learned representations (*information*). We introduced varying levels of heterogeneous ET→MC weight duplication, comparing layer utilization and information measures to a homogeneous control condition in which network topology was matched but all ET→MC connections had identical weights. We also altered the distribution of ET→MC quantization weights based on the statistics of a calibration set, with the aim of demonstrating that utilization regularization can be traded off for enhanced resolution in a target range. These objectives yielded four experimental conditions, contingent on two key independent variables: (1) ET→MC weight distribution; and (2) duplication factor (MC:ET ratio; 4, 8, 16, or 32).

In the *homogeneous* condition, the number of MCs was identical to the number of ETs (8); one of 10 weights was picked out of a $\frac{1}{x}$ distribution covering the input range $[.01, 1]$ (with respect to the chosen LIF parameters for MCs) and applied to every ET→MC synapse. The MC duplication factor varied among the levels 4x, 8x, 16x, and 32x, and samples could be drawn from the concentration or the saturation datasets. Thus, this design yielded a total of 80 homogeneous scenarios, each corresponding to an independent simulation run.

In the three *heterogeneous* conditions, duplicated MCs were arranged in columns defined by having a common ET afferent; ET→MC weights within each column were varied. Each heterogeneous condition yielded 8 independent simulation runs (4 duplication levels, 2 datasets), for a total of 24 runs. In the *uniform* variant, ET→MC weights covered the input range $[.01, 1]$. In the *scaled* variant, they covered the range $[min(x_i), max(x_i)]$ across calibration set observations $x$. The *adaptive* variant was similar to the *scaled* one, except the densities of the ET→MC weights were selected to approximate the exact distribution of calibration observations for each sensor, with a certain degree of smoothing. Specifically, within each duplication factor level $n$, each sensor's range was partitioned into $n$ segments. Optimal partition points were determined using one-dimensional clustering (Jenks natural breaks optimization) [19]; the same number of weights was assigned to each segment despite potential differences in width, yielding a variable density distribution reflecting the statistics of the held out set used for training and calibration.

To quantify spiking layer activity *regularization*, we computed the standard deviation of the percentage of units that spiked in response to each test sample, separately for the MC and GC layers within each network model. Minimizing the variance in spiking layer utilization across input samples stabilizes representations and plasticity mechanisms, maintaining EPL network activity in the sensitive, well-tuned midrange and also ensuring that concentration differences are not mistaken for analyte properties and exploited for discrimination. To quantify the *information* encoded in the learned EPL representations, we fed MC and GC responses to a cross validated, hyperparameter optimized support vector machine (SVM) classifier with a radial basis function (RBF) kernel (for details, see *Procedure*). Classification accuracy was used as a measure of whether the transformation (i.e., analog input vectors to MC and GC spike phases) resulted in more (or less) easily separable representations, using the ET responses as a baseline (as those were equivalent to the normalized analog input vectors).

**Procedure**

We performed a total of 104 simulation runs spanning four quantization conditions in varying architectural configurations (see *Experimental Design*). Independent simulations were performed with a consistent random seed across the relevant libraries (*random*, *numpy*, and *pytorch*), ensuring reproducibility of the processing steps described below. All weights and states were simulated with float32 precision on an Intel i9-14900k CPU; however, network components abide by neuromorphic design principles and can be readily deployed on Intel Loihi processors [5].

*Sampling*. At the beginning of each simulation run, 128 vectors were sampled from either the *concentration* or *saturation* dataset; samples were identical across independent instances of the pipeline. Within each instance, eight-fold cross validation was performed; an initial shuffling of the sample indices was followed by their partitioning into non-overlapping splits, each consisting of 16 training samples (two per analyte class, consistent with two-shot learning) and 112 test samples (14 per analyte class).

*Calibration*. In the *scaled* and *adaptive* heterogeneous quantization conditions, we introduced a pre-training calibration procedure. Within each cross validation fold, the network was exposed to the 14 training set samples with synaptic learning turned off, and the range and density of ET→MC weights within each organizational column were updated accordingly (see *Experimental Design*).

*Model Training and Testing*. During training, the model was continuously exposed to each sample for 250 steps (corresponding to five oscillatory cycles), followed by a 50 step rinse cycle in which the network received an all-zeros vector as its input. MC→GC synaptic weights were updated at the end of every cycle in accordance with a Hebbian STDP rule (for parameterization, see *Network Architecture*). The same exposure and rinse durations were used in the test trials, with learning turned off. Response vectors (ET, MC, and GC) from the final (fifth) exposure cycle for each sample were logged and used in subsequent statistical and classification analyses.

*Classification*. To gauge the quality of the mapping learned by our EPL model, ET, MC, and GC spike-phase response vectors (*processed responses*) were fed, respectively, to three independent RBF kernel support vector machine (SVM) classifiers (using *scikit-learn* [33]). The following procedure was performed independently within each model-level cross validation fold (train-test split): first, hyperparameter optimization was performed ($10^{-7} < C < 10^9$, $10^{-8} < \gamma < 10^{10}$, with logarithmically spaced intervals) using half (56) of the processed responses; then, SVMs were trained and tested on the remaining 56 samples. Stratified 4-fold cross validation was used in both steps, with the smaller segment used as the internal training set (consistent with few-shot learning). Classification predictions were aggregated across these folds to yield confusion matrices containing prediction outcomes for 21 samples per analyte class (168 in total). Classification accuracy scores were calculated as the proportion of correct predictions out of the total number of processed responses in the test set. Thus, within each independent simulation run (corresponding to a single experimental condition) and layer (ET, MC, GC), eight classification scores were obtained and subsequently used to compute 95% confidence intervals around the score mean.

**Software**

Simulations were performed in Python 3.11 with *pytorch* 2.1.2 and *sapicore* 0.3.3. Sapicore is an open source spiking neural network modeling framework developed by the authors to streamline the design and testing of neuromorphic algorithms on CPU/GPU. Sampling and classification were performed using *scikit-learn* 1.3.2. Statistical analyses were performed in R 4.1.2 using the packages *lme4*, *emmeans*, and *multcomp*.

**Data Availability**

Source code for *Sapicore* is freely available on GitHub (https://github.com/cplab/sapicore). Our synthetic data, primary results, SVG figure elements, and statistics can be reproduced by running master scripts (Python and R) provided in a separate repository accompanying this manuscript (https://github.com/cplab/sapinet_regularization).

**Acknowledgments**

This work was supported by:

- NSF NCS CBET-2123862 (TAC).
- NSF EFRI BRAID EFMA-2223811 (TAC).
- Intel INRC Faculty Grant (TAC).
- Teledyne FLIR (TAC).
- The Eric and Wendy Schmidt AI in Science Postdoctoral Fellowship, a Schmidt Futures Program (RM).

## Author Contributions

RM implemented the network, designed the experiments, generated the data and figures, and performed the analyses. AB and TAC outlined preliminary models of heterogeneity-based data regularization. RM and TAC developed the algorithm and wrote the manuscript. KRM and TAC wrote the supplementary proof. RM and ME co-developed *Sapicore*.

## Additional Information

*Competing Interests*. TAC and AB are listed as inventors on US patent US20220198245A1 and European patent EP3953868A4, both pending, on neuromorphic methods for rapid online learning and signal restoration. RM, KRM, and ME declare no competing interests.

## Supplementary Methods

### Data Synthesis

*Affinities.* To produce balanced correlated affinities inducing a nested similarity structure with three levels, we started with a matrix whose rows were identical to $G_{1,:}$ (below), then recursively swapped adjacent submatrices of decreasing sizes (powers of 2). The process yielded the following symmetric affinity matrix, whose sparsity was subsequently adjusted until a Hoyer measure of 0.4 was reached (see *Methods*):

$$G = \begin{pmatrix} 1 & .95 & .8 & .75 & .4 & .35 & .2 & .15 \\ .95 & 1 & .75 & .8 & .35 & .4 & .15 & .2 \\ .8 & .75 & 1 & .95 & .2 & .15 & .4 & .35 \\ .75 & .8 & .95 & 1 & .15 & .2 & .35 & .4 \\ .4 & .35 & .2 & .15 & 1 & .95 & .8 & .75 \\ .35 & .4 & .15 & .2 & .95 & 1 & .75 & .8 \\ .2 & .15 & .4 & .35 & .8 & .75 & 1 & .95 \\ .15 & .2 & .35 & .4 & .75 & .8 & .95 & 1 \end{pmatrix}$$

*Sampling.* Sensor response vectors sampled from the *concentration* dataset can be written as $OSN(x) = Gx + \varepsilon$, where $\varepsilon$ is a Gaussian noise term $\sim N(0, \sigma_i)$ and $x$ is a column vector with one nonzero element $x_i \sim \exp(\lambda) + c$. The parameters $\sigma$, $\lambda$, and $c$ were varied across, but not within, analyte classes $i$ as follows: $c = (1,2,4,8,16,32,64,128)$, $\lambda = (10,10,1,1,1,1,1,1)$, $\sigma = (.01, .01, .01, .01, .01, .02, .05, .1)$, yielding the samples depicted in Figs. 2D, S1B. In the *saturation* dataset, the total output emitted by the sensors was $OSN(x) = \frac{1}{1 + e^{-8(Gx - .5)}} + \varepsilon$, with the noise component parameterized $\varepsilon \sim N(0, .02)$ across analyte classes; other parameters were identical to the *concentration* set, resulting in the samples displayed in Figs. 2H, S1D. The resulting classification problem was nontrivial due to higher levels of either variance or additive noise in some analyte classes, but still retained a well-defined hierarchical similarity structure.

This parameterization yielded the following unsorted raw and normalized responses:

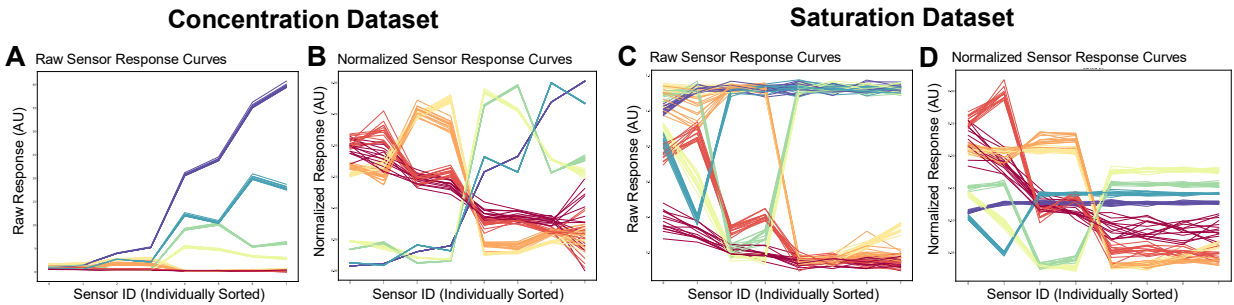

**Supplementary Figure S1**. **Unsorted raw and normalized sensor responses to test samples from two synthetically generated datasets**. Samples are color coded by analyte class. (**A**) Dataset with balanced sensor affinities and systematic differences in synthetic analyte concentration. (**B**) A sigmoid transformed variant of the first dataset, with imbalanced response magnitudes and systematic differences in sensor saturation across analytes.

## Supplementary Data

### Statistical Analyses

The line plots below are an alternative presentation of the data in Fig. 5:

### Spiking Layer Regularization by Quantization Method

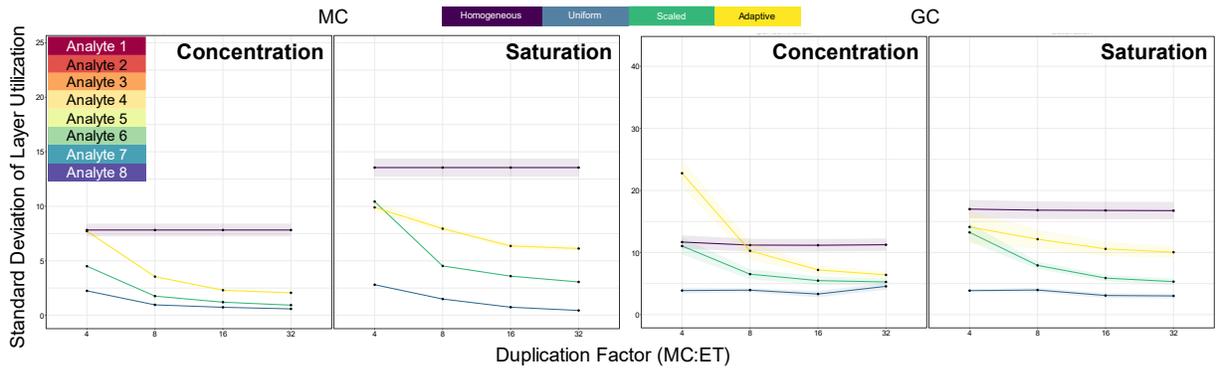

**Supplementary Figure S2**. **Line plots depicting spiking layer utilization regularization**. Regularization is defined as the standard deviation of the percentage of MCs or GCs that spiked in response to sensory input (lower is better). It is presented as a function of the duplication factor. Ribbons represent the standard error of the mean over eight cross validation folds. Colors denote the weight distribution strategy (see Fig. 5 for details).

### Analyte Decoding Performance by Quantization Method

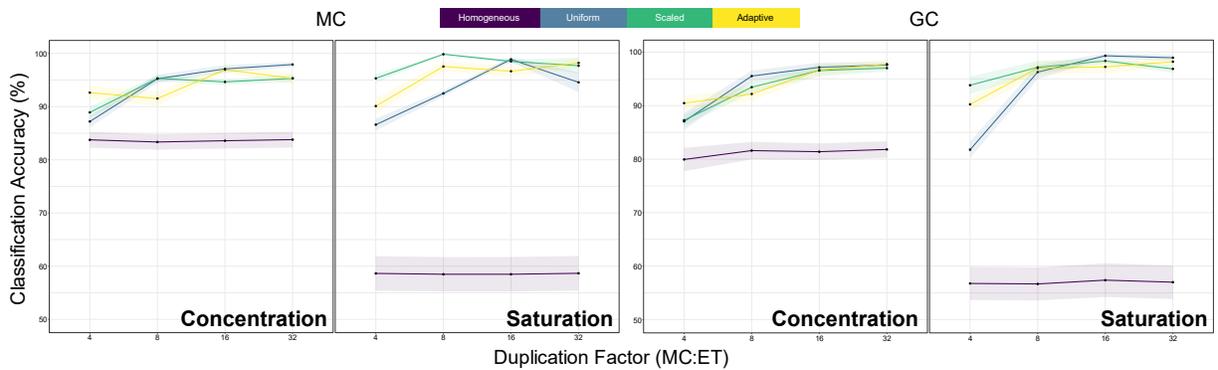

**Supplementary Figure S3**. **Analyte classification accuracy from spike phase responses**. ET, MC, and GC response vectors were used as inputs to cross validated, hyperparameter optimized RBF SVMs. Analyte classification accuracies are presented as a function of the duplication factor. Ribbons represent the standard error of the mean over eight cross validation folds. Colors denote the weight distribution strategy (see Fig. 5 for details).

**Analysis of the Regularizing Effects of Gain Diversification**

In the present neuromorphic model, non-spiking external tufted (ET) cells deliver excitation onto spiking mitral cells (MC; Fig. 1). This is the quantization step, at which analog sensory inputs are transduced into spike timing-based representations. Each MC receives excitatory input from exactly one ET cell and is activated (spikes) if and only if this input exceeds its intrinsic spike threshold. In contrast, ET cells excite all of the MCs within their column, the number of which is determined by the duplication factor. Sensory inputs evoke activity in ET cells as follows:

$$A = [a_0 \ a_1 \ a_2 \ldots a_{n-1}]$$

where $n$ is the total number of ET cells (corresponding to the number of columns) and $a_i$ is the activity of the $i^{th}$ ET cell. As discussed in the Results, ET cell activity is normalized such that the total activation across the ET cell population sums to a constant $c$.

$$\sum_{i=0}^{n-1} a_i = c$$

For visualization purposes, we depict the activity of 8 ET cells (corresponding to 8 sensor-associated columns) as a line graph, sorted in descending order of the amplitude of their activation (Supplementary Fig. S4A; *black*). We will refer to this descending function as $f$. We can approximate the value of the area $c$ under the function $f$ as the Riemann sum of the areas of a number of discrete rectangles with finite widths (Supplementary Fig. S4A; *blue rectangles*):

$$Area(f) = \int_0^n f(x)dx \approx \sum_{i'=0}^{n'} f(x_{i'})\Delta x_{i'}$$

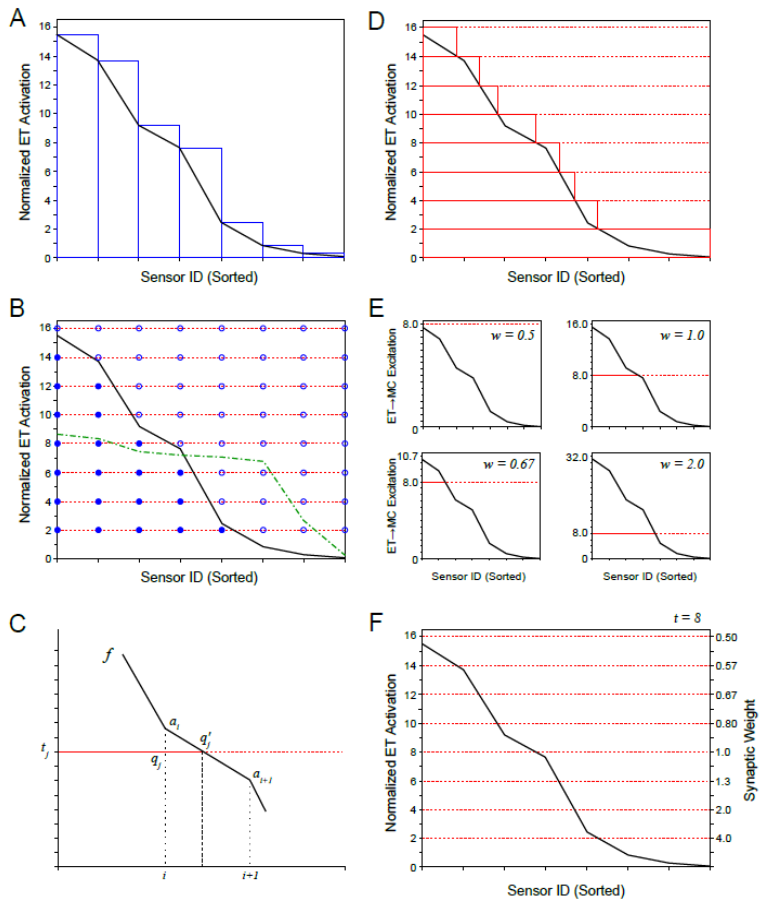

**Supplementary Figure S4**. (**A**) Normalized ET response profile of an analyte across 8 sensors (abscissa; cf. Fig. 2B, 2F), with Riemann subdivisions (*blue rectangles*). (**B**) The same response profile as in A, annotated with eight different, uniformly spaced, MC spike thresholds (*red dotted lines*; implies a duplication factor of 8). Each of the 8 sensor-associated ET cells activates one MC at each threshold, for a total of 64 MCs (*blue circles*). MCs below/to the left of the ET response profile (*black curve*) produce action potentials (*solid blue circles*). A normalized ET response profile from a different analyte (*green broken curve*) activates approximately the same number of MCs (*blue circles beneath green broken curve*), illustrating the regularization of total MC activity irrespective of input distribution. (**C**) Determination of the fraction of the MC population at a certain threshold that fires (*solid portion of red threshold line*), for a segment of the function $f$. (**D**) Same response profile as in A, annotated with eight subdivisions for calculation of the inverted Riemann sum (across thresholds rather than sensors). (**E**) Effect of heterogeneous ET→MC synaptic weights with a uniform MC threshold of 8. Greater weights generate larger numbers of activated MCs in the same way as lower MC thresholds do. (**F**) A $\frac{1}{x}$ distribution of synaptic weights yields the same outcome as a uniform distribution of MC thresholds.

Substituting the literal indices as $x_i, = i$, for which we know the value of $f(i) = a_i$, yields:

$$Area(f) \approx \sum_{i=0}^{n-1} f(i)\Delta x_i = \sum_{i=0}^{n-1} a_i \Delta x_i$$

Because the width of each rectangle is exactly one, and the sum of all ET cell activities is a constant:

$$Area(f) \approx \sum_{i=0}^{n-1} a_i$$

The approximation tends to the precise area under the curve as $n$ increases.

*Regularization of MC Spiking Activity via Heterogeneous MC Spike Thresholds*

We first will demonstrate that heterogeneous MC spike *thresholds* will regularize the number of MCs activated by any sensory input distribution, given a constant area $c$ under the function $f$ of ET cell activation levels after normalization, and assuming that all ET→MC synaptic weights are identical and that the MC thresholds are uniformly distributed and span the entire codomain of $f$.

Assume that each ET cell projects to $m$ MCs within its column (Fig. 1). Each of these MCs is assigned a different spike threshold, uniformly distributed between zero and some fixed upper bound $u$. The list of $m$ thresholds will be the same in all columns. Hence, for $n$ columns, there are $n$ MCs exhibiting each threshold, and the total number of MCs in the network is $nm$. The assumption that the MC thresholds span the entire codomain of $f$ means that $u \geq a_0$, though the general finding is robust to minor deviations from this assumption.

Consider $ET_i$, the $i^{th}$ ET cell with activity $a_i$. Let $ET_i$ synapse onto $MC_{i,j}$ where $i$ is the index of the presynaptic ET cell. Let $MC_{i,j}$ have threshold $t_j$. An $MC_{i,j}$ is considered to be active (spiking) if and only if $a_i \geq t_j$. A uniform distribution of eight MC thresholds is depicted in Supplementary Fig. S4B (*dotted red lines*), based on a duplication factor of 8.

We define the total number of MCs that are firing as $Q$. Since the the total number of MCs in the system is $nm$, the fraction of MCs that are firing is $\frac{Q}{nm}$. We want to demonstrate that this fraction is approximately constant for all normalized sensory inputs irrespective of their distribution (Fig. 2B, 2F). As MCs will spike if and only if their threshold is below the activity function $f$, it also is true, given that ET cell activation levels are sorted by amplitude, that MCs spike if and only if their thresholds are to the left of $f$ (Supplementary Fig. S4B, *closed blue circles*). Heuristically, a different normalized input distribution, such as one drawn from the saturation dataset (Fig. 2F), still activates an approximately equal number of MCs (Supplementary Fig. S4B, *total count of the open and closed blue circles beneath the function g depicted by broken green lines*).

To demonstrate this formally, Supplementary Fig. S4C depicts a section of function $f$. We define $q_j'$ as the X coordinate of the point where $f$ intersects the MC threshold $t_j$. Because only MCs with thresholds to the left of $f$ will spike, $MC_{1,j}, MC_{2,j}, \ldots MC_{i,j}$ will spike and $MC_{i+1,j}, \ldots MC_{n,j}$ will not. That is, $i$ out of $n$ MCs at threshold $t_j$ will fire. We therefore can define $q_j$ as the (integer) number of spiking MCs at a given threshold $t_j$. In the above case, $q_j = i$. Because $|q_j' - q_j| < 1$, the value of $q_j$ is always a strong (close) approximation of the value of $q_j'$. For large values of $n$, the fraction $\frac{q_j}{n} \approx \frac{q_j'}{n}$ represents the fraction of MCs that are spiking. That is, approximately $q'_j$ MCs at each threshold $t_j$ spike.

For a function like $f$ that is non-increasing, with domain $[0, n]$ and codomain $[0, a_0]$, the Riemann sum can be used on either axis to estimate the area under the curve. That is, the regular Riemann sum and the inverted Riemann sum approximate the same area:

$$Area(f) = \int_0^n f(x)dx \approx \sum_{i'=0}^{n'} f(x_{i'})\Delta x_{i'} \approx \sum_{j=0}^{m'} f^{-1}(y_j)\Delta y_j$$

We therefore can use the inverted Riemann sum to approximate the area under the function $f$ using the evenly spaced MC thresholds $t_j$ as uniform subdivisions (Fig. S2D). Substituting $t_j$ for $y_j$, and defining a constant $\Delta y$ as the difference between each pair of thresholds:

$$Area(f) \approx \sum_{j=0}^{m-1} f^{-1}(t_j)\Delta y$$

Because $f^{-1}(t_j) = q_j' \approx q_j$:

$$Area(f) \approx \sum_{j=0}^{m-1} q_j \Delta y = \Delta y \sum_{j=0}^{m-1} q_j$$

Because $\sum_{j=0}^{m-1} q_j$ is the sum of the number of MCs of all thresholds that are spiking, and is therefore equal to the total number of MCs that are spiking:

$$Q = \sum_{j=0}^{m-1} q_j$$

$$Area(f) \approx Q\Delta y$$

Because $Area(f) = c$, for constant $n$:

$$c \approx Q\Delta y$$
$$Q \approx \frac{c}{\Delta y}$$
$$\frac{Q}{nm} \approx \frac{c}{nm\Delta y}$$

Because the thresholds of the $m$ MCs for each ET cell are spread evenly from 0 to $u$, yielding $\Delta y = \frac{u}{m}$, this finally resolves to:

$$\frac{Q}{nm} \approx \frac{c}{nu}$$

where $c, n, u$ are all constants. That is, given the above assumptions and a constant $n, c, u$, the fraction of MCs that are firing approximates a constant (compare counts of *blue circles* to the left of black function $f$ and green function $g$ in Supplementary Fig. S4B). This finding extends to random MC thresholds drawn from a uniform distribution, with or without equivalence across columns (not shown). Using the Monte Carlo method, as the duplication factor (equivalent to $m$ if thresholds are heterogeneous and drawn from a uniform distribution) increases, the fraction of activated MCs will approximate the fraction of the total area under the curve:

$$\frac{\#Firing\ MCs}{\#Total\ MCs} \approx \frac{Area(f)}{Sample\ Area}$$

$$\frac{Q}{nm} \approx \frac{c}{nu}$$

That is, given a sufficient duplication factor with MC thresholds drawn from a uniform distribution spanning the entire codomain of $f$, the fraction of activated MCs in response to normalized sensor input of any distribution will approximate a constant.

*Regularization of MC Spiking Activity via Heterogeneous ET→MC Synaptic Weights*

We now consider the effects of heterogeneous ET→MC synaptic weights rather than heterogeneous MC thresholds. Here, each ET cell synapses onto $m$ MCs exhibiting identical thresholds $t$, with the synaptic weights $w_j$ from $ET_i$ to $MC_{i,j}$ being uniformly spaced from the range $[l, u]$. $MC_{i,j}$ spikes if and only if $a_i w_j \geq t$. As above, we define the number of MCs with a given incoming weight $w_j$ to be $q_j$ (which will equal the number of columns). To depict the resulting generation of spikes in MCs with identical thresholds $t = 8$, we group all MCs by their incoming synaptic weights $w_j$, and plot four of these weight groups for comparison (Supplementary Fig. S4E). The synaptic weight $w_j$ determines the ordinate scale for each plot. As above, the fraction of the horizontal threshold line to the left of the function $f$ (Supplementary Fig. S4E, *solid red lines*) represents the number of MCs with the incoming weight $w_j$ that are spiking.

To rescale each of the functions of Supplementary Fig. S4E onto a common ordinate scale (depicting the activation of the presynaptic ET cell), we divide each threshold $t$ by the weight $w_j$ for each plot to generate an *equivalent threshold* $t_j' = t/w_j$.

For the distribution of equivalent thresholds $t_j'$ to be uniform, as depicted above (Supplementary Fig. S4B), the weights $w_j$ must be distributed as $\frac{1}{x}$ Supplementary Fig. S4F). That is, with the above assumptions, normalized input, and given a uniform MC spiking threshold $t$, a $\frac{1}{x}$ distribution of heterogeneous ET→MC weights ensures that the fraction of MCs that are firing approximates a constant irrespective of the input distribution.

**Descriptive and Inferential Statistics**

| Layer | Dataset | Condition | Duplication | Accuracy Mean | Accuracy SD |
|---|---|---|---|---|---|
| GC | Concentration | Homogeneous | 4 | 79.94047619 | 0.947051414 |
| GC | Concentration | Homogeneous | 8 | 81.59970238 | 1.543177004 |
| GC | Concentration | Homogeneous | 16 | 81.39136905 | 1.56257086 |
| GC | Concentration | Homogeneous | 32 | 81.81547619 | 1.097564816 |
| GC | Concentration | Uniform | 4 | 87.05357143 | 4.519219834 |
| GC | Concentration | Uniform | 8 | 95.53571429 | 2.967674892 |
| GC | Concentration | Uniform | 16 | 97.17261905 | 2.459380079 |
| GC | Concentration | Uniform | 32 | 97.61904762 | 1.558699216 |
| GC | Concentration | Scaled | 4 | 87.27678571 | 2.88003689 |
| GC | Concentration | Scaled | 8 | 93.45238095 | 2.863513306 |
| GC | Concentration | Scaled | 16 | 96.57738095 | 3.129552078 |
| GC | Concentration | Scaled | 32 | 97.02380952 | 1.683587574 |
| GC | Concentration | Adaptive | 4 | 90.47619048 | 3.883737159 |
| GC | Concentration | Adaptive | 8 | 92.1875 | 3.563447715 |
| GC | Concentration | Adaptive | 16 | 96.65178571 | 2.754268634 |
| GC | Concentration | Adaptive | 32 | 97.76785714 | 2.540368919 |
| GC | Saturation | Homogeneous | 4 | 56.75595238 | 0.972887227 |
| GC | Saturation | Homogeneous | 8 | 56.66666667 | 0.765589517 |
| GC | Saturation | Homogeneous | 16 | 57.3735119 | 1.090364369 |
| GC | Saturation | Homogeneous | 32 | 56.99404762 | 0.829681015 |
| GC | Saturation | Uniform | 4 | 81.77083333 | 4.338505874 |
| GC | Saturation | Uniform | 8 | 96.2797619 | 3.720663241 |
| GC | Saturation | Uniform | 16 | 99.33035714 | 1.250101227 |
| GC | Saturation | Uniform | 32 | 98.95833333 | 1.645578442 |
| GC | Saturation | Scaled | 4 | 93.82440476 | 4.631915198 |
| GC | Saturation | Scaled | 8 | 97.17261905 | 3.49623218 |
| GC | Saturation | Scaled | 16 | 98.36309524 | 1.413970166 |
| GC | Saturation | Scaled | 32 | 96.875 | 2.005973829 |
| GC | Saturation | Adaptive | 4 | 90.25297619 | 3.817189089 |
| GC | Saturation | Adaptive | 8 | 97.02380952 | 3.45619072 |
| GC | Saturation | Adaptive | 16 | 97.24702381 | 3.290195542 |
| GC | Saturation | Adaptive | 32 | 98.21428571 | 2.227177016 |
| MC | Concentration | Homogeneous | 4 | 83.75744048 | 0.914110702 |
| MC | Concentration | Homogeneous | 8 | 83.34821429 | 1.070365118 |
| MC | Concentration | Homogeneous | 16 | 83.58630952 | 0.937786777 |
| MC | Concentration | Homogeneous | 32 | 83.79464286 | 0.935625353 |
| MC | Concentration | Uniform | 4 | 87.20238095 | 3.149703942 |
| MC | Concentration | Uniform | 8 | 95.23809524 | 2.110488715 |
| MC | Concentration | Uniform | 16 | 97.09821429 | 2.0481109 |
| MC | Concentration | Uniform | 32 | 97.91666667 | 1.311840869 |
| MC | Concentration | Scaled | 4 | 88.91369048 | 3.05072486 |
| MC | Concentration | Scaled | 8 | 95.3125 | 1.72626378 |
| MC | Concentration | Scaled | 16 | 94.64285714 | 1.590840726 |
| MC | Concentration | Scaled | 32 | 95.3125 | 2.02324672 |
| MC | Concentration | Adaptive | 4 | 92.63392857 | 3.527758807 |
| MC | Concentration | Adaptive | 8 | 91.51785714 | 3.827947583 |

| Layer | Dataset | Condition | Duplication | Mean | SD |
|---|---|---|---|---|---|
| MC | Concentration | Adaptive | 16 | 96.94940476 | 1.696689621 |
| MC | Concentration | Adaptive | 32 | 95.3125 | 2.975127361 |
| MC | Saturation | Homogeneous | 4 | 58.63839286 | 1.150448454 |
| MC | Saturation | Homogeneous | 8 | 58.48214286 | 1.365160112 |
| MC | Saturation | Homogeneous | 16 | 58.48214286 | 1.164251978 |
| MC | Saturation | Homogeneous | 32 | 58.66071429 | 1.258273559 |
| MC | Saturation | Uniform | 4 | 86.60714286 | 3.229054004 |
| MC | Saturation | Uniform | 8 | 92.48511905 | 1.620394473 |
| MC | Saturation | Uniform | 16 | 98.88392857 | 1.328613945 |
| MC | Saturation | Uniform | 32 | 94.56845238 | 5.222587395 |
| MC | Saturation | Scaled | 4 | 95.3125 | 2.34750098 |
| MC | Saturation | Scaled | 8 | 99.85119048 | 0.275541696 |
| MC | Saturation | Scaled | 16 | 98.51190476 | 1.147171561 |
| MC | Saturation | Scaled | 32 | 97.69345238 | 4.202199077 |
| MC | Saturation | Adaptive | 4 | 90.10416667 | 4.171977907 |
| MC | Saturation | Adaptive | 8 | 97.54464286 | 2.612807579 |
| MC | Saturation | Adaptive | 16 | 96.65178571 | 2.897558189 |
| MC | Saturation | Adaptive | 32 | 98.21428571 | 2.718429824 |

**Supplementary Table S1**. Analyte classification accuracy means and standard deviations as a function of spiking layer, dataset, condition, and duplication factor.

| Contrast | Duplication | Layer | Estimate | SE | df | t-ratio | p-value |
|---|---|---|---|---|---|---|---|
| Homogeneous - Uniform | 4 | GC | -16.0639881 | 0.860475203 | 441 | -18.66874031 | 0 |
| Homogeneous - Scaled | 4 | GC | -22.20238095 | 0.860475203 | 441 | -25.8024646 | 0 |
| Homogeneous - Adaptive | 4 | GC | -22.01636905 | 0.860475203 | 441 | -25.58629114 | 0 |
| Uniform - Scaled | 4 | GC | -6.138392857 | 0.860475203 | 441 | -7.133724295 | 2.55418E-12 |
| Uniform - Adaptive | 4 | GC | -5.952380952 | 0.860475203 | 441 | -6.917550831 | 7.57473E-11 |
| Scaled - Adaptive | 4 | GC | 0.186011905 | 0.860475203 | 441 | 0.216173463 | 0.996428898 |
| Homogeneous - Uniform | 8 | GC | -26.77455357 | 0.860475203 | 441 | -31.11600833 | 0 |
| Homogeneous - Scaled | 8 | GC | -26.17931548 | 0.860475203 | 441 | -30.42425325 | 0 |
| Homogeneous - Adaptive | 8 | GC | -25.47247024 | 0.860475203 | 441 | -29.60279409 | 0 |
| Uniform - Scaled | 8 | GC | 0.595238095 | 0.860475203 | 441 | 0.691755083 | 0.900230925 |
| Uniform - Adaptive | 8 | GC | 1.302083333 | 0.860475203 | 441 | 1.513214244 | 0.430396697 |
| Scaled - Adaptive | 8 | GC | 0.706845238 | 0.860475203 | 441 | 0.821459161 | 0.844312547 |
| Homogeneous - Uniform | 16 | GC | -28.86904762 | 0.860475203 | 441 | -33.55012153 | 0 |
| Homogeneous - Scaled | 16 | GC | -28.08779762 | 0.860475203 | 441 | -32.64219298 | 0 |
| Homogeneous - Adaptive | 16 | GC | -27.56696429 | 0.860475203 | 441 | -32.03690729 | 0 |
| Uniform - Scaled | 16 | GC | 0.78125 | 0.860475203 | 441 | 0.907928547 | 0.800633048 |
| Uniform - Adaptive | 16 | GC | 1.302083333 | 0.860475203 | 441 | 1.513214244 | 0.430396697 |
| Scaled - Adaptive | 16 | GC | 0.520833333 | 0.860475203 | 441 | 0.605285698 | 0.930375032 |
| Homogeneous - Uniform | 32 | GC | -28.88392857 | 0.860475203 | 441 | -33.56741541 | 0 |
| Homogeneous - Scaled | 32 | GC | -27.54464286 | 0.860475203 | 441 | -32.01096647 | 0 |
| Homogeneous - Adaptive | 32 | GC | -28.58630952 | 0.860475203 | 441 | -33.22153787 | 0 |
| Uniform - Scaled | 32 | GC | 1.339285714 | 0.860475203 | 441 | 1.556448937 | 0.404785457 |
| Uniform - Adaptive | 32 | GC | 0.297619048 | 0.860475203 | 441 | 0.345877542 | 0.98576716 |
| Scaled - Adaptive | 32 | GC | -1.041666667 | 0.860475203 | 441 | -1.210571395 | 0.62044913 |
| Homogeneous - Uniform | 4 | MC | -15.70684524 | 0.860475203 | 441 | -18.25368726 | 0 |
| Homogeneous - Scaled | 4 | MC | -20.91517857 | 0.860475203 | 441 | -24.30654423 | 0 |
| Homogeneous - Adaptive | 4 | MC | -20.17113095 | 0.860475203 | 441 | -23.44185038 | 0 |
| Uniform - Scaled | 4 | MC | -5.208333333 | 0.860475203 | 441 | -6.052856977 | 1.82241E-08 |
| Uniform - Adaptive | 4 | MC | -4.464285714 | 0.860475203 | 441 | -5.188163123 | 1.93604E-06 |
| Scaled - Adaptive | 4 | MC | 0.744047619 | 0.860475203 | 441 | 0.864693854 | 0.823056835 |
| Homogeneous - Uniform | 8 | MC | -22.94642857 | 0.860475203 | 441 | -26.66715845 | 0 |
| Homogeneous - Scaled | 8 | MC | -26.66666667 | 0.860475203 | 441 | -30.99062772 | 0 |
| Homogeneous - Adaptive | 8 | MC | -23.61607143 | 0.860475203 | 441 | -27.44538292 | 0 |
| Uniform - Scaled | 8 | MC | -3.720238095 | 0.860475203 | 441 | -4.323469269 | 0.000111582 |
| Uniform - Adaptive | 8 | MC | -0.669642857 | 0.860475203 | 441 | -0.778224469 | 0.864309676 |
| Scaled - Adaptive | 8 | MC | 3.050595238 | 0.860475203 | 441 | 3.545244801 | 0.002443481 |
| Homogeneous - Uniform | 16 | MC | -26.95684524 | 0.860475203 | 441 | -31.32785833 | 0 |
| Homogeneous - Scaled | 16 | MC | -25.54315476 | 0.860475203 | 441 | -29.68494 | 0 |
| Homogeneous - Adaptive | 16 | MC | -25.76636905 | 0.860475203 | 441 | -29.94434816 | 0 |
| Uniform - Scaled | 16 | MC | 1.413690476 | 0.860475203 | 441 | 1.642918322 | 0.355621578 |
| Uniform - Adaptive | 16 | MC | 1.19047619 | 0.860475203 | 441 | 1.383510166 | 0.510361743 |
| Scaled - Adaptive | 16 | MC | -0.223214286 | 0.860475203 | 441 | -0.259408156 | 0.993876629 |
| Homogeneous - Uniform | 32 | MC | -25.01488095 | 0.860475203 | 441 | -29.07100737 | 0 |
| Homogeneous - Scaled | 32 | MC | -25.27529762 | 0.860475203 | 441 | -29.37365022 | 0 |
| Homogeneous - Adaptive | 32 | MC | -25.53571429 | 0.860475203 | 441 | -29.67629307 | 0 |
| Uniform - Scaled | 32 | MC | -0.260416667 | 0.860475203 | 441 | -0.302642849 | 0.990364532 |
| Uniform - Adaptive | 32 | MC | -0.520833333 | 0.860475203 | 441 | -0.605285698 | 0.930375032 |
| Scaled - Adaptive | 32 | MC | -0.260416667 | 0.860475203 | 441 | -0.302642849 | 0.990364532 |

**Supplementary Table S2**. Analyte classification accuracy general linear contrast estimates, standard errors (SE), degrees of freedom (df), t-ratios, and p-values.

| Model Term | df1 | df2 | F-ratio | p-value |
|---|---|---|---|---|
| Layer | 1 | 441 | 5.805 | 0.016394082 |
| Condition | 3 | 441 | 3298.391 | 1.4056E-301 |
| Duplication | 3 | 441 | 179.135 | 6.09289E-76 |
| Dataset | 1 | 441 | 555.771 | 4.1112E-80 |
| Layer:Condition | 3 | 441 | 6.572 | 0.000235806 |
| Layer:Duplication | 3 | 441 | 5.944 | 0.000557259 |
| Layer:Dataset | 1 | 441 | 0.042 | 0.838225874 |
| Condition:Duplication | 9 | 441 | 28.307 | 9.38911E-39 |
| Condition:Dataset | 3 | 441 | 937.419 | 6.6972E-191 |
| Duplication:Dataset | 3 | 441 | 3.08 | 0.027316623 |
| Layer:Condition:Duplication | 9 | 441 | 1.656 | 0.097428339 |
| Layer:Condition:Dataset | 3 | 441 | 1.498 | 0.214322582 |
| Layer:Duplication:Dataset | 3 | 441 | 0.047 | 0.986298344 |
| Condition:Duplication:Dataset | 9 | 441 | 7.022 | 1.65477E-09 |
| Layer:Condition:Duplication:Dataset | 9 | 441 | 2.719 | 0.004295063 |

**Supplementary Table S3**. Analyte classification accuracy GLM effects, degrees of freedom, F-ratios, and p-values.

| Layer | Dataset | Condition | Duplication | Regularization Mean | Regularization SD |
|---|---|---|---|---|---|
| GC | Concentration | Homogeneous | 4 | 11.6864796 | 3.651219251 |
| GC | Concentration | Homogeneous | 8 | 11.19987165 | 3.23680018 |
| GC | Concentration | Homogeneous | 16 | 11.17477839 | 3.074444542 |
| GC | Concentration | Homogeneous | 32 | 11.25678322 | 3.1179467 |
| GC | Concentration | Uniform | 4 | 3.879449763 | 1.321959286 |
| GC | Concentration | Uniform | 8 | 3.942542236 | 0.916368666 |
| GC | Concentration | Uniform | 16 | 3.313069463 | 1.508523858 |
| GC | Concentration | Uniform | 32 | 4.555809921 | 1.553856109 |
| GC | Concentration | Scaled | 4 | 11.05426512 | 3.902302178 |
| GC | Concentration | Scaled | 8 | 6.517569274 | 2.27642365 |
| GC | Concentration | Scaled | 16 | 5.478909897 | 1.98343533 |
| GC | Concentration | Scaled | 32 | 5.263088826 | 1.550317777 |
| GC | Concentration | Adaptive | 4 | 22.77290058 | 6.131622727 |
| GC | Concentration | Adaptive | 8 | 10.27835366 | 4.272892392 |
| GC | Concentration | Adaptive | 16 | 7.201965751 | 1.918027423 |
| GC | Concentration | Adaptive | 32 | 6.400802718 | 1.675925721 |
| GC | Saturation | Homogeneous | 4 | 16.9932849 | 0.840534859 |
| GC | Saturation | Homogeneous | 8 | 16.82709804 | 0.710959865 |
| GC | Saturation | Homogeneous | 16 | 16.78360853 | 0.678076916 |
| GC | Saturation | Homogeneous | 32 | 16.74944532 | 0.763515635 |
| GC | Saturation | Uniform | 4 | 3.86770553 | 0.676724556 |
| GC | Saturation | Uniform | 8 | 3.961989606 | 1.071082407 |
| GC | Saturation | Uniform | 16 | 3.08050323 | 1.108177677 |
| GC | Saturation | Uniform | 32 | 3.013742946 | 1.135873255 |
| GC | Saturation | Scaled | 4 | 13.24707783 | 4.220504491 |
| GC | Saturation | Scaled | 8 | 7.928806939 | 1.765918795 |
| GC | Saturation | Scaled | 16 | 5.887620437 | 1.421716172 |
| GC | Saturation | Scaled | 32 | 5.315451537 | 1.538495658 |
| GC | Saturation | Adaptive | 4 | 14.12113731 | 7.698267568 |
| GC | Saturation | Adaptive | 8 | 12.15457229 | 4.242918981 |
| GC | Saturation | Adaptive | 16 | 10.5922558 | 3.266749874 |
| GC | Saturation | Adaptive | 32 | 10.05655451 | 2.680447304 |
| MC | Concentration | Homogeneous | 4 | 7.824148097 | 0.074457813 |
| MC | Concentration | Homogeneous | 8 | 7.824145469 | 0.074398295 |
| MC | Concentration | Homogeneous | 16 | 7.824148097 | 0.074457813 |
| MC | Concentration | Homogeneous | 32 | 7.824148097 | 0.074457813 |
| MC | Concentration | Uniform | 4 | 2.244967548 | 0.030675823 |
| MC | Concentration | Uniform | 8 | 0.960367407 | 0.02110347 |
| MC | Concentration | Uniform | 16 | 0.737411469 | 0.010411331 |
| MC | Concentration | Uniform | 32 | 0.59154363 | 0.006409456 |
| MC | Concentration | Scaled | 4 | 4.503799784 | 0.078281626 |
| MC | Concentration | Scaled | 8 | 1.761200722 | 0.062733902 |
| MC | Concentration | Scaled | 16 | 1.205629684 | 0.061771946 |
| MC | Concentration | Scaled | 32 | 0.933814636 | 0.07090784 |
| MC | Concentration | Adaptive | 4 | 7.719650827 | 0.416014399 |
| MC | Concentration | Adaptive | 8 | 3.551177311 | 0.230092824 |
| MC | Concentration | Adaptive | 16 | 2.301491171 | 0.198062336 |
| MC | Concentration | Adaptive | 32 | 2.065476444 | 0.131475482 |
| MC | Saturation | Homogeneous | 4 | 13.55384004 | 0.016201584 |
| MC | Saturation | Homogeneous | 8 | 13.55384004 | 0.016201584 |
| MC | Saturation | Homogeneous | 16 | 13.55384004 | 0.016201584 |
| MC | Saturation | Homogeneous | 32 | 13.55384004 | 0.016201584 |
| MC | Saturation | Uniform | 4 | 2.807856339 | 0.050181508 |
| MC | Saturation | Uniform | 8 | 1.494912709 | 0.040691747 |
| MC | Saturation | Uniform | 16 | 0.744845508 | 0.022611407 |
| MC | Saturation | Uniform | 32 | 0.447346685 | 0.017835972 |
| MC | Saturation | Scaled | 4 | 10.43286881 | 0.274201495 |
| MC | Saturation | Scaled | 8 | 4.52793637 | 0.123755745 |
| MC | Saturation | Scaled | 16 | 3.594095616 | 0.053574454 |
| MC | Saturation | Scaled | 32 | 3.062618808 | 0.061640575 |
| MC | Saturation | Adaptive | 4 | 9.898485804 | 1.492510947 |
| MC | Saturation | Adaptive | 8 | 7.949284134 | 0.750913344 |
| MC | Saturation | Adaptive | 16 | 6.361624816 | 0.830858088 |
| MC | Saturation | Adaptive | 32 | 6.127168534 | 0.64767331 |

**Supplementary Table S4**. Utilization regularization (standard deviation of active % of MC and GC layers) means and standard deviations as a function of spiking layer, dataset, condition, and duplication factor.

| Contrast | Duplication | Layer | Estimate | SE | df | t-ratio | p-value |
|---|---|---|---|---|---|---|---|
| Homogeneous - Uniform | 4 | GC | 10.46630461 | 0.716147152 | 441 | 14.61474026 | 0 |
| Homogeneous - Scaled | 4 | GC | 2.189210775 | 0.716147152 | 441 | 3.056928692 | 0.012644709 |
| Homogeneous - Adaptive | 4 | GC | -4.107136693 | 0.716147152 | 441 | -5.735045772 | 1.08456E-07 |
| Uniform - Scaled | 4 | GC | -8.27709383 | 0.716147152 | 441 | -11.55781157 | 0 |
| Uniform - Adaptive | 4 | GC | -14.5734413 | 0.716147152 | 441 | -20.34978603 | 0 |
| Scaled - Adaptive | 4 | GC | -6.296347469 | 0.716147152 | 441 | -8.791974464 | 0 |
| Homogeneous - Uniform | 8 | GC | 10.06121892 | 0.716147152 | 441 | 14.04909438 | 0 |
| Homogeneous - Scaled | 8 | GC | 6.790296736 | 0.716147152 | 441 | 9.48170599 | 0 |
| Homogeneous - Adaptive | 8 | GC | 2.797021865 | 0.716147152 | 441 | 3.905652433 | 0.000626949 |
| Uniform - Scaled | 8 | GC | -3.270922185 | 0.716147152 | 441 | -4.56738839 | 3.79243E-05 |
| Uniform - Adaptive | 8 | GC | -7.264197057 | 0.716147152 | 441 | -10.14344195 | 0 |
| Scaled - Adaptive | 8 | GC | -3.993274871 | 0.716147152 | 441 | -5.576053557 | 2.5693E-07 |
| Homogeneous - Uniform | 16 | GC | 10.78240712 | 0.716147152 | 441 | 15.0561335 | 0 |
| Homogeneous - Scaled | 16 | GC | 8.295928297 | 0.716147152 | 441 | 11.58411128 | 0 |
| Homogeneous - Adaptive | 16 | GC | 5.082082691 | 0.716147152 | 441 | 7.096422404 | 9.19897E-12 |
| Uniform - Scaled | 16 | GC | -2.48647882 | 0.716147152 | 441 | -3.472022216 | 0.00317331 |
| Uniform - Adaptive | 16 | GC | -5.700324426 | 0.716147152 | 441 | -7.959711093 | 0 |
| Scaled - Adaptive | 16 | GC | -3.213845606 | 0.716147152 | 441 | -4.487688877 | 5.42657E-05 |
| Homogeneous - Uniform | 32 | GC | 10.21833783 | 0.716147152 | 441 | 14.26848911 | 0 |
| Homogeneous - Scaled | 32 | GC | 8.713844085 | 0.716147152 | 441 | 12.16767262 | 0 |
| Homogeneous - Adaptive | 32 | GC | 5.774435652 | 0.716147152 | 441 | 8.063197123 | 0 |
| Uniform - Scaled | 32 | GC | -1.504493748 | 0.716147152 | 441 | -2.100816494 | 0.1544181 |
| Uniform - Adaptive | 32 | GC | -4.443902181 | 0.716147152 | 441 | -6.205291988 | 7.52837E-09 |
| Scaled - Adaptive | 32 | GC | -2.939408433 | 0.716147152 | 441 | -4.104475494 | 0.000281197 |
| Homogeneous - Uniform | 4 | MC | 8.162582124 | 0.716147152 | 441 | 11.39791188 | 0 |
| Homogeneous - Scaled | 4 | MC | 3.220659771 | 0.716147152 | 441 | 4.497203912 | 5.20084E-05 |
| Homogeneous - Adaptive | 4 | MC | 1.879925753 | 0.716147152 | 441 | 2.625055128 | 0.044243049 |
| Uniform - Scaled | 4 | MC | -4.941922353 | 0.716147152 | 441 | -6.900707965 | 8.67574E-11 |
| Uniform - Adaptive | 4 | MC | -6.282656371 | 0.716147152 | 441 | -8.772856749 | 0 |
| Scaled - Adaptive | 4 | MC | -1.340734019 | 0.716147152 | 441 | -1.872148783 | 0.24157166 |
| Homogeneous - Uniform | 8 | MC | 9.461352695 | 0.716147152 | 441 | 13.21146454 | 0 |
| Homogeneous - Scaled | 8 | MC | 7.544424208 | 0.716147152 | 441 | 10.53474023 | 0 |
| Homogeneous - Adaptive | 8 | MC | 4.938762031 | 0.716147152 | 441 | 6.896295015 | 8.98387E-11 |
| Uniform - Scaled | 8 | MC | -1.916928488 | 0.716147152 | 441 | -2.676724307 | 0.03848173 |
| Uniform - Adaptive | 8 | MC | -4.522590664 | 0.716147152 | 441 | -6.315169522 | 3.93219E-09 |
| Scaled - Adaptive | 8 | MC | -2.605662176 | 0.716147152 | 441 | -3.638445215 | 0.001739077 |
| Homogeneous - Uniform | 16 | MC | 9.947865579 | 0.716147152 | 441 | 13.89081219 | 0 |
| Homogeneous - Scaled | 16 | MC | 8.289131418 | 0.716147152 | 441 | 11.57462038 | 0 |
| Homogeneous - Adaptive | 16 | MC | 6.357436074 | 0.716147152 | 441 | 8.877276214 | 0 |
| Uniform - Scaled | 16 | MC | -1.658734161 | 0.716147152 | 441 | -2.316191802 | 0.095888886 |
| Uniform - Adaptive | 16 | MC | -3.590429505 | 0.716147152 | 441 | -5.013535971 | 4.61822E-06 |
| Scaled - Adaptive | 16 | MC | -1.931695344 | 0.716147152 | 441 | -2.697344169 | 0.036368778 |
| Homogeneous - Uniform | 32 | MC | 10.16954891 | 0.716147152 | 441 | 14.20036216 | 0 |
| Homogeneous - Scaled | 32 | MC | 8.690777346 | 0.716147152 | 441 | 12.13546312 | 0 |
| Homogeneous - Adaptive | 32 | MC | 6.592671578 | 0.716147152 | 441 | 9.205749914 | 0 |
| Uniform - Scaled | 32 | MC | -1.478771565 | 0.716147152 | 441 | -2.064899039 | 0.166330192 |
| Uniform - Adaptive | 32 | MC | -3.576877332 | 0.716147152 | 441 | -4.994612245 | 5.0668E-06 |
| Scaled - Adaptive | 32 | MC | -2.098105767 | 0.716147152 | 441 | -2.929713206 | 0.018658942 |

**Supplementary Table S5**. Utilization regularization (standard deviation of active % of MC and GC layers) general linear contrast estimates, standard errors (SE), degrees of freedom (df), t-ratios, and p-values.

| Model Term | df1 | df2 | F-ratio | p-value |
| --- | --- | --- | --- | --- |
| Layer | 1 | 441 | 476.183 | 3.96306E-72 |
| Condition | 3 | 441 | 559.527 | 6.8404E-150 |
| Duplication | 3 | 441 | 91.858 | 3.40888E-46 |
| Dataset | 1 | 441 | 177.808 | 2.56168E-34 |
| Layer:Condition | 3 | 441 | 17.085 | 1.63219E-10 |
| Layer:Duplication | 3 | 441 | 3.314 | 0.019955889 |
| Layer:Dataset | 1 | 441 | 22.519 | 2.81649E-06 |
| Condition:Duplication | 9 | 441 | 21.523 | 3.1216E-30 |
| Condition:Dataset | 3 | 441 | 44.078 | 6.25657E-25 |
| Duplication:Dataset | 3 | 441 | 2.041 | 0.107374497 |
| Layer:Condition:Duplication | 9 | 441 | 3.765 | 0.000138808 |
| Layer:Condition:Dataset | 3 | 441 | 4.683 | 0.00312121 |
| Layer:Duplication:Dataset | 3 | 441 | 4.212 | 0.005929607 |
| Condition:Duplication:Dataset | 9 | 441 | 8.261 | 2.14031E-11 |
| Layer:Condition:Duplication:Dataset | 9 | 441 | 2.721 | 0.004274631 |

**Supplementary Table S6**. Utilization regularization (standard deviation of active % of MC and GC layers) GLM effects, degrees of freedom, F-ratios, and p-values.